\documentclass[a4paper,structabstract]{aa}
\pdfoutput=1
\usepackage{aas_macros}
\usepackage{natbib}
\usepackage{txfonts}
\usepackage{array}
\usepackage{wasysym} % atronomical symbols
\usepackage{ifpdf} % help while compiling with PDFLaTeX
\usepackage{multirow}

\ifpdf                                  % si le compilateur est PDFLaTeX :
 \usepackage[pdftex]{graphicx}          % pour l'insertion d'images (graphicx) et la gestion des couleurs (color)
 \usepackage[pdftex]{thumbpdf}          % pour creer des vignettes pdf
 \usepackage[pdftex,
       bookmarks = true,               % Signets
       bookmarksnumbered = true,       % Signets numérotés
       pdfpagemode = UseNone,             % Signets/vignettes fermé à l'ouverture
       pdfstartview = FitH,            % La page prend toute la largeur
       pdfpagelayout = SinglePage,     % Vue par page
       backref,                        % pour les liens dans la biblio
       hyperfootnotes = true,          % pour les notes de bas de page
       colorlinks = true,              % Liens en couleur
       urlcolor = blue,                % Couleur des liens externes
       linkcolor = red,                % couleur des liens internes
       anchorcolor = black,            % couleur des ancres du document
       citecolor = blue,              % couleur des réf bibliographique
       pdfborder = {0 0 0}             % Style de bordure : ici, pas de bordure
       ]{hyperref}                     % Utilisation de HyperTeX
 \hypersetup{% information sur le PDF créé
       baseurl = {http://www.aanda.org}, % url où se trouve le document
       pdftitle = {Analysis of the CoRoT-2 system: A young spotted star and its inflated giant planet}, % titre
       pdfsubject = {CoRoT-2b}, % sujet
       pdfauthor = {Tristan GUILLOT, Mathieu HAVEL}, % auteur
       pdfkeywords = {extrasolar giant planets -- planet formation}, % mots-clés
       pdfcreator = {Kile, LaTeX, hyperref}, % createur (logiciel)
       pdfproducer = {PDFLaTeX}, % producteur
}
 \DeclareGraphicsExtensions{.pdf, .png, .jpg} % pour les extensions des images incluses
\else                                   % sinon, avec le compilateur LaTeX :
 \usepackage{graphicx}                  % pour l'insertion d'images
 \usepackage[
        colorlinks = true,              % Liens en couleur
        urlcolor = blue,                % Couleur des liens externes
        linkcolor = red,                % couleur des liens internes
        anchorcolor = black,            % couleur des ancres du document
        citecolor = green               % couleur des réf bibliographique
        ]{hyperref}                     % package permettant de mettre en forme les liens hypertextes
 \DeclareGraphicsExtensions{.eps, .ps, .pst}
\fi

\def\modif{}

\def\wig#1{\mathrel{\hbox{\hbox to 0pt{%
          \lower.6ex\hbox{$\sim$}\hss}\raise.4ex\hbox{$#1$}}}}

\def\mjup{$\rm M_J$}
\def\rjup{$\rm R_J$}

\def\teff{T_{\rm eff}}
\def\teffs{T_{\rm eff,\,spectro}}
\def\teffm{\overline{T}_{\rm eff}}
\def\teffsm{\overline{T}_{\rm eff,\,spectro}}

\def\kapv{\kappa_{\rm v}}
\def\kapth{\kappa_{\rm th}}
\def\tint{T_{\rm int}}

\def\teq{T_{\rm eq}}

\def\aap{{\it Astron. \& Astrophys}}

\begin{document}

\title{An analysis of the CoRoT-2 system: A young spotted star and its inflated giant planet}
\titlerunning{CoRoT-2}
\authorrunning{Guillot \& Havel}

\author{Tristan Guillot
  \and
  Mathieu Havel
  }

\offprints{T. Guillot}

\institute{Universit\'e de Nice-Sophia Antipolis, CNRS UMR 6202, Observatoire de la C\^ote d'Azur, B.P. 4229, 06304 Nice Cedex 4, France\\
           \email{tristan.guillot, mathieu.havel@oca.eu}}

\date{Submitted to A\&A: 26 May 2010; Revised: 4 Sept. 2010; Accepted: 13 Sept. 2010}

\abstract
{CoRoT-2b is one of the most anomalously large exoplanet known. Given its high mass, its large radius cannot be explained by standard evolution models. Interestingly, the planet's parent
star is an active, rapidly rotating solar-like star with with spots covering a large fraction (7-20\%) of its visible surface.}
{We attempt to constrain the properties of the star-planet system and
  understand whether the planet's inferred large size may be caused a systematic error in the inferred parameters, and if not, how it can be explained.
}
{We combine stellar and planetary evolution codes based on all available spectroscopic and photometric data to obtain self-consistent constraints on the system parameters.
}
{We find no systematic error in the stellar modeling (including spots and stellar activity) that would cause a $\sim 10\%$ reduction in size of the star and thus the planet.
Two classes of solutions are found: the usual main-sequence solution for the star yields for the planet a mass of $3.67\pm 0.13\,$\mjup, a radius of $1.55\pm 0.03$\,\rjup\ for an age that is
at least $130\,$Ma and should be younger than $500\,$Ma given the star's rapid rotation and significant activity. We identify another class of solutions on the pre-main sequence, for which
the planet's mass is $3.45\pm 0.27\,$\mjup\ and its radius is $1.50\pm 0.06$\,\rjup\ for an age of $30$ to $40$\,Ma. These extremely young solutions provide the simplest explanation
of the planet's size that can then be matched by a simple contraction from an initially hot, expanded state, if the atmospheric opacities are larger by a factor of $\sim 3$
than usually assumed for solar composition atmospheres. Other solutions imply that the present inflated radius of CoRoT-2b is transient and the result of an event
that occurred less than $20\,$Ma ago, i.e., a giant impact with another Jupiter-mass planet, or interactions with another object in the system that caused a significant rise in the eccentricity
followed by the rapid circularization of its orbit.
}
{Additional observations of CoRoT-2 that could help us to understand this
  system include searches for an infrared excess, 
a debris disk, and additional companions. The determination of a complete infrared lightcurve including both the primary and secondary transits would also be extremely valuable to
constrain the planet's atmospheric properties and determine the planet-to-star
    radius ratio in a manner less vulnerable to systematic errors caused by stellar activity.}

% \abstract
% {The exoplanet CoRoT-2b is presently the most anomalously
%   large planet known. Its large mass ($\sim 3$\,\mjup) and large
%   radius ($\sim 1.5$\,\rjup) cannot be explained by standard evolution
%   models including stellar irradiation but not extra heat sources or
%   modified input physics. Interestingly, the planet's parent star is
%   an active star with a large fraction (7 to 20\%) of spots.}
% {Aims: We use both stellar and planetary models to determine whether the
%   presence of starspots may affect the parameters infered for the
%   system, and evaluate how the planet's radius may be explained.}
% {Methods: Using different evolution codes ...}
% {Results:}
% {}


\maketitle
%
%________________________________________________________________

\section{Introduction}

CoRoT-2b is the second transiting planet discovered by the space
mission CoRoT \citep{Alonso+08, Bouchy+08}. It is
noteworthy because of its relatively large mass and large size. As
such, it belongs to the class of anomalously large extrasolar giant
planets, i.e. planets that are larger than predicted by standard
theoretical models for the evolution of an irradiated,
solar-composition gas giant \citep{Guillot+06}.

Anomalously large planets require additional heat sources
\citep[e.g.][]{BLM01, GS02, Baraffe+03, Guillot+06, BHBH07} to explain their
radii. It is now clear that these objects are common, representing 
about half of the presently known transiting planets
\citep{Guillot08,MFJ09}. However the properties of CoRoT-2b are probably the most difficult to explain:
it has {\modif one of} the largest radius anomalies (defined as the difference between
observed and modeled for a solar-composition irradiated planet), and
yet is massive, meaning that modifying its evolution to slow its
contraction is difficult and requires significant deviations from the
standard models.

{\em Secondary} transits of CoRoT-2b have been detected by several
instruments providing puzzling results: Measured brightness
temperatures have been found to vary from $1325\pm 180\,$K at 8\,$\mu$m to $1805\pm
70\,$K at 4.5\,$\mu$m \citep{Gillon+10,Snellen+10}, and up to $2170\pm
50\,$K in the visible \citep{Snellen+10,Alonso+10}, significantly higher than
the zero-albedo equilibrium temperature of the planet $1530\pm
140$\,K.

The star CoRoT-2 is also remarkable because of its unusual variability
and spot coverage: analyses of the CoRoT lightcurve for this system
--spanning 152 days of nearly-continuous observations-- led to a
precise determination of the star's rotation period, $4.52\pm
0.14\,$days \citep{Lanza+09}, and estimates of the spot coverage
{\modif that range between 7\% and 20\%, for a spot contrast of between $\sim 0.3$ and
  $0.7$ \citep{Lanza+09, HCWS10, Silva-Valio+10}. The spot coverage
(i.e. the fraction of the stellar surface that is occupied by
starspots) may even locally reach 37\% for a contrast of 0.7 (similar
to the average value for sunspots) at the latitudes where the
planetary eclipses occur \citep{HCWS10}. }

Could the peculiarities of the star account for the unusually large inferred
size of the planet or do we require additional heat sources in the
planet? We address this question by first revisiting the star's
evolution by accounting for the presence of spots (Sect.~2), then applying
these results to planetary evolution models (Sect.~3).

\section{The evolution of a spotted star: Constraints on CoRoT-2}

\subsection{Constraints derived from spectroscopic and transit photometry analyses}

The star  CoRoT-2 has been identified as a G7 dwarf of solar composition by
spectroscopic analyses. Table~\ref{tab:obs} lists its known physical
properties inferred from spectroscopy, radial velocimetry, and the analysis of its
transit lightcurve.

\begin{table}
\caption{Observational constraints on the stellar parameters}
\label{tab:obs}
\begin{tabular}{lcl}
\hline
\multicolumn{3}{c}{\textit{Spectroscopy}}\\
$T_{\rm eff, H\alpha}$  & $5450\pm 120$\,K & \citet{Bouchy+08} \\
$T_{\rm eff, [Fe]}$ & $5625\pm 120$\,K & \citet{Alonso+08} \\
$T_{\rm eff}$ & $5608\pm 37$\,K & \citet{AmmlervonEiff+09}\\
$\log g$ & $4.3\pm 0.2\rm\,cm\,s^{-2}$ & \citet{Alonso+08}  \\
$\log g$ & $4.71\pm 0.2\rm\,cm\,s^{-2}$ & \citet{AmmlervonEiff+09} \\
$\rm [M/H]$ & $0.0\pm 0.1$ & \citet{Alonso+08} \\
& &\\
\multicolumn{3}{c}{\textit{Photometry and RV}}\\
$m_{\rm V}$ & 12.57 & \citet{Alonso+08} \\
$P_\star$ & $4.522\pm 0.024$ days & \citet{Lanza+09} \\
$P_{\rm orb}$  & $1.7429935\pm 10^{-6}$\,days & \citet{Gillon+10}\\
%======== Modif Mathieu ============
$k$ & ${\modif 0.1711\pm 0.0011}$ & {\rm This work} \\
$k$ & ${\modif 0.1658\pm 0.0004}$ & \citet{Gillon+10}\\
$k$ & ${\modif 0.1720\pm 0.0010}$ & \citet{Czesla+09}\\
$k$ & ${\modif 0.1667\pm 0.0006}$ & \citet{Alonso+08}\\
%========= End Modif ==========
$i$ & $88.08^{+0.18}_{-0.16}$ deg & \citet{Gillon+10}\\
$\rho_\star$ & $1.870\pm 0.026\rm\,g\,cm^{-3}$ & \citet{Alonso+08} \\
$\rho_\star$ & $1.814^{+0.050}_{-0.045}\rm\,g\,cm^{-3}$ & \citet{Gillon+10} \\
$K$ & $603\pm 18\rm\,m\,s^{-1}$ & \citet{Gillon+10} \\
$a/R_\star$ & $6.64\pm 0.03$ & \citet{Gillon+10}\\
\hline
\end{tabular}
\end{table}

The effective temperature of CoRoT-2 differs slightly between
measurements: \citet{Bouchy+08} report a low value from H$_\alpha$
bands and relatively low signal-to-noise HARPS spectra, and a high
value, also from HARPS but using Fe spectral lines. \citet{AmmlervonEiff+09}
essentially confirm this last value with Fe lines and UVES, but with a
smaller error bar (which does not however include systematic effects).

The spectral determination of the star's gravity is, as commonly found for stars, quite
uncertain. Measurements inferred from HARPS and UVES data differ slightly,
there being one $\sigma$ error bars that barely overlap.

As is typical of a star with a transiting planet, the most stringent
constraint is that on the stellar density. For CoRoT-2, \citet{Alonso+08} are
able to determine the duration of the transit so precisely that the
stellar density is constrained to within only 1.4\%.

Given these measurements, \citet{Alonso+08} infer planetary parameters,
$M_{\rm p}=3.31\pm 0.16$\,\mjup\ and $R_{\rm p}=1.465\pm 0.029$\,
\rjup. This implies that CoRoT-2b is extremely inflated even relative 
to other large transiting planets. Before we attempt to model it, we
estimate the accuracy to which 
the stellar parameters can really be derived. For example, this
initial estimate by \citeauthor{Alonso+08} assumes a circular orbit
and no influence by spots on the photometric determination.

A subsequent analysis of the CoRoT-2 lightcurves \citep{Czesla+09} shows
that the presence of spots during the transits affects the photometry
and in particular the depth $k^2$ of the transits: when the planet
transits, it occasionally occults star spots, thereby blocking a smaller fraction of
the stellar light. On average, the transits appear less deep than for
a star without spots, implying that the planetary radius is
underestimated when neglecting the effect of spots.

The effect of a non-zero eccentricity is to modify the relation
between transit duration and stellar density in a non-trivial way relative to a circular orbit
\citep[e.g.][]{Tingley05}. \citet{Gillon+10} refined the analysis of
\citeauthor{Alonso+08} using constraints on the eccentricity and
argument of the periastron from the radial velocity data within a
Markov Chain Monte Carlo approach. They found
that solutions with a slight eccentricity ($e\sim 0.015$) are more
likely and thus inferred a lower stellar density $\rho_\star$
and a slightly smaller value of $k$.

Most of the parameters used for this work are based on the analysis of
\citet{Gillon+10} (see Table~\ref{tab:obs}). However, to account for the effect of
spots on the transit depths, and because the \citet{Czesla+09} study
did not allow for the possibility of an eccentric planet, we choose to
derive a probable value of $k$ that accounts for both spots and a
non-circular orbit by a simple proportionality relation between the
different studies: $ k = k_{\rm
  Czesla} / k_{\rm Alonso} \times k_{\rm Gillon}$. The error bar in $k$ is
calculated to be the quadratic mean between the values of $k$ found by
the \citeauthor{Czesla+09} and \citeauthor{Gillon+10} studies.

\subsection{Evolution models}

Stellar evolution models are needed to derive the star's and planet's
masses and radii. In most of this work, we use a grid of quasi-static
evolutions for stars with masses between 0.6 and 1.3 ${\rm M}_{\odot}$
($\rm \Delta M_\star = 0.005 M_\odot$) calculated with the CESAM
evolution code \citep{ML08}. The grid has been calibrated with respect
to the Sun, which is most accurately described  ($< 10^{-4}$ relative precision) by mass
fractions of hydrogen $X_\odot = 0.7065$, helium $Y_\odot = 0.2740$,
and heavy elements $Z_\odot = 0.0195$, based on the actual $Z/X =
0.0245$ ratio of \citet{Grevesse+93}, and spectroscopic parameters
$\rm T_{eff,\odot} = 5778$ K, and $\rm L_\odot = 3.846\times 10^{26}$ W
for an age of 4.57\,Ga.  A standard mixing-length approach without
overshooting is used in the energy transport equation. Our calibrated
Sun has a mixing length parameter $\alpha_\odot
= 2.052$. The atmospheric boundary is calculated using a $T(\tau)$
relation derived from MARCS models \citep{Gustafsson+08}, and we
consider the microscopic diffusion of chemical elements in the
radiative zone (therefore the atmospheric metallicity varies as a 
function of time). We chose the abundances of
\citet{Grevesse+93} because the seismological
constraints are not met when using other, more recent abundances.

For comparison, we also use the grid of models calculated by
\citet{BCAH98} (hereafter BCAH98) for solar-composition stars. Those
models assume a non-grey atmosphere, and the convection is treated in
the context of the mixing-length theory with no core-overshooting and
no diffusion. The tables assume that $X = 0.716$, $Y = 0.282$,
$Z = 0.020$, and $\alpha = 1.9$. For a solar model
($1\,M_\odot$, $1\,R_\odot$), the tables yield an effective
temperature $\rm T_{eff,\odot} = 5797$\,K and luminosity $\rm L_\odot
= 3.801\times 10^{26}$\,W for an assumed age of $4.61$\,Ga.

Finally, results are also compared to the so-called $Y^2$ stellar evolution
tracks (hereafter YY) for solar composition stars
\citep{Demarque+04}. These models include convection-overshoot and
diffusion.

\subsection{The effect of spots and activity}

By definition, the effective temperature of a star $\teff$ is linked to
its total luminosity $L$ and radius $R$ by the well-known
relation
\begin{equation}
L=4\pi R^2 \sigma \teff^4
\label{eq:teff}
\end{equation}
However, in the presence of spots and activity, this relation has
to be revised, because the correspondence between the effective temperature
derived from a spectroscopic analysis (which we refer to as $\teffs$) and that obtained from
Eq.~\ref{eq:teff} no longer holds.

Across the face of the Sun, spots are indicative of high
magnetic activity: although the spots block a fraction ($\sim 0.1\%$)
of the starlight, their appearance is connected to a global {\it
  increase} in the total solar irradiance by about 0.1\%, from $\sim
1365.5$ to $1366.5\rm\,W\,m^{-2}$ due to the presence of bright
faculae \citep[e.g.][]{FL04, KS08}. This
implies that the flux emitted in the visible and most importantly in
the UV increases. Secular variations in the total solar irradiance
based on models and a $\sim 300$ year record of the solar activity
also amount to about $0.1\%$ \citep{SF00}.

For stars with activity levels comparable to that of the Sun or lower,
this level of uncertainty is much smaller than that attainable
from spectroscopic measurements. In the case of CoRoT-2, the area
covered by starspots is 70 to 200 times larger than for the active Sun
and thus can potentially affect what may be inferred about the star
properties, i.e. its luminosity, radius and mass.

In Table~\ref{tab:obs}, the effective temperature of CoRoT-2 was
determined from either the H$-\alpha$ line at 656.3\,nm or fits
in the visible (450 to 740\,nm). For comparison, between minimum and
maximum of activity, the Sun increases its relative flux $\Delta
F_\lambda / F_\lambda$ by about 0.1\%, 0.05\%, and 0.07\% at
wavelengths in the range of 400-500\,nm, 500-700\,nm, and 700-800\,nm,
respectively \citep{Krivova+06}. Variations in the brightness
temperatures at these wavelengths (proportional to $\sigma T^4$) are 4
times smaller. {\it If} we assume that these number
indeed scale with the surface area covered by spots (a big {\it if},
arguably), the potential mismatch between the spectroscopically
inferred $\teffs$ and the true $\teff$ amounts to up to 5\%.

To see how the results may be affected by
stellar activity, we first compare standard evolution models to
evolution models calculated with CESAM and a modified atmospheric boundary
condition
\begin{equation}
L=(1-\chi_{\rm s}) 4\pi R^2 \sigma \teffs^4,
\label{eq:teffs}
\end{equation}
{\modif where $\chi_{\rm s}$ is a factor that accounts for emission from only 
part of the surface of the star. For purely black spots
with no faculae it corresponds to the surface fraction of spots.}

\begin{figure}
\centerline{\resizebox{8.5cm}{!}{\includegraphics[angle=-90]{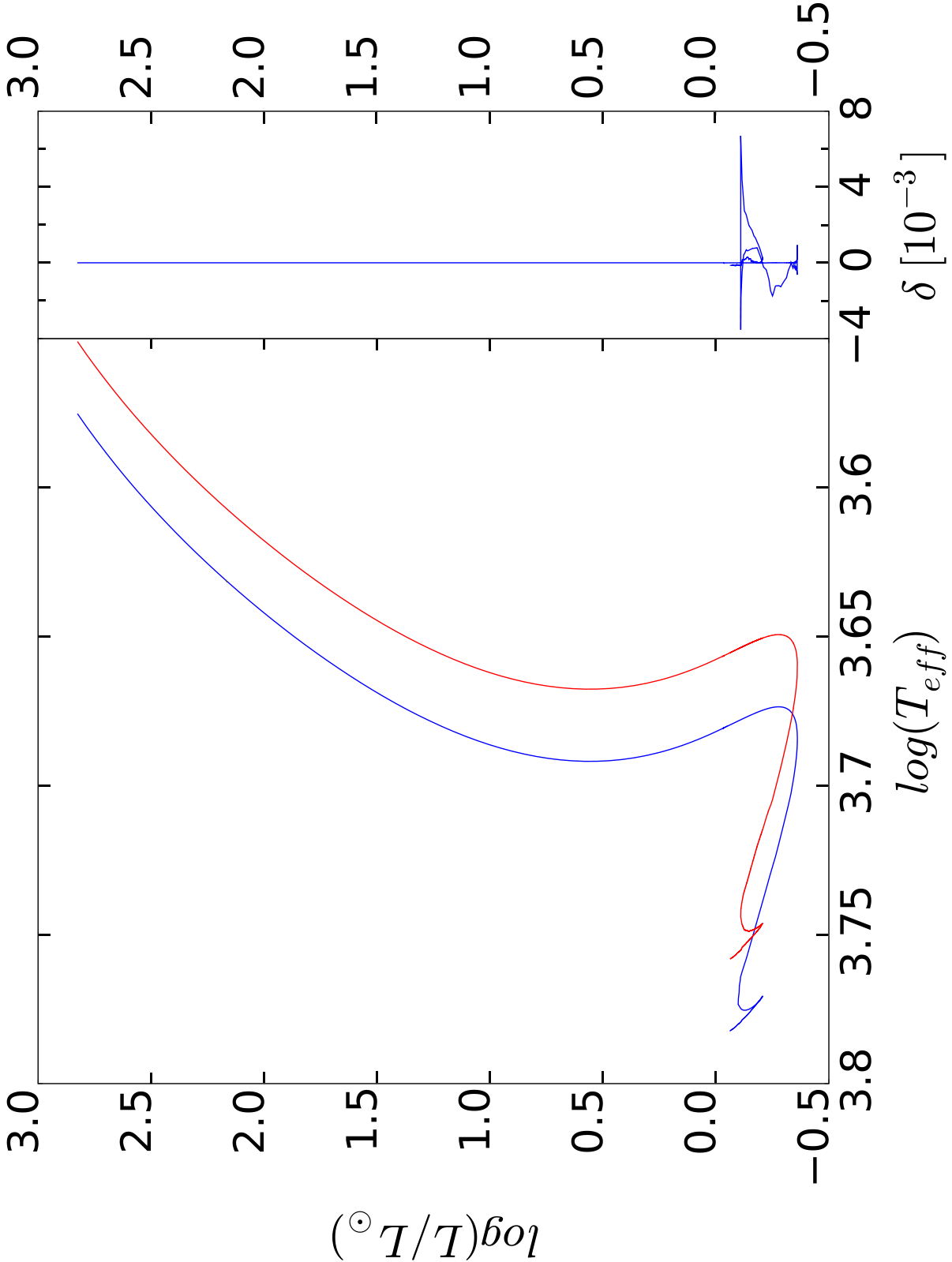}}}
\caption{\textit{Left panel:} Hertzsprung-Russell evolution tracks for
  0.97\,M$_\odot$, Z = 0.02, $\alpha$ = 1.9 for a standard evolution
  model (red line), and when modifying the atmospheric boundary
  condition to account for the presence of $\chi_{\rm s}=20\%$ of
  {\modif dark} spots on the
  stellar surface (blue line). \textit{Right panel:} differences in
  effective temperature for a given luminosity between {\modif the} model with no
  spot, and {\modif the} model with spots {\modif after} the effective temperature has
  been shifted by a constant factor $(1-\chi_{\rm s})^{1/4}$.}
\label{fig:hr_comp}
\end{figure}

Figure~\ref{fig:hr_comp} shows the result for a given luminosity and
age, a star with spots simply has a higher ``spectroscopic'' effective
temperature $\teffs$ as inferred from spectroscopic measurements than
a star without spots. The evolution is simply displaced to the left of
the HR diagram by an almost constant ratio $(1-\chi_{\rm s})^{1/4}$ in
$\teff$. Quantitatively, the mean deviation amounts to $5\times 10^{-5}$ on
$\modif \teff$, with a standard deviation $\sigma=4.6\times 10^{-4}\modif\teff$
and a maximum deviation $6.7\times 10^{-3}\modif\teff$.  The small
departures from this constant displacement are due to slight
modifications of the atmospheric properties (opacities) with 
temperature, but they can safely be neglected in this study.

When considering a star's evolution, that a star has spots is
therefore equivalent to an added uncertainty in the measurement of its
effective temperature. {\modif Activity also has the same consequence
  because it implies that both the present spectroscopically-determined
  effective temperature and the present luminosity may differ from
  their value averaged over one stellar magnetic cycle.} A larger
error bar in $\teff$ {\modif may be used} as a proxy for the {\modif
  added uncertainty due to} activity and starspots. {\modif Although
  this uncertainty may be either positive (if the luminosity of CoRoT-2 is
  lower than its mean value and/or the contribution of faculae is
  larger than that of spots) or negative (if CoRoT-2 is more luminous
  than average and/or the spots are dark) we choose to only study the
  latter possibility. As shown in the following section, only
  lower $\teff$ values can yield smaller radii for the star and the planet
  and hence help to solve the inflation puzzle.}

\subsection{Constraints on the star's physical parameters}

\begin{figure}
\centerline{\resizebox{8.5cm}{!}{\includegraphics{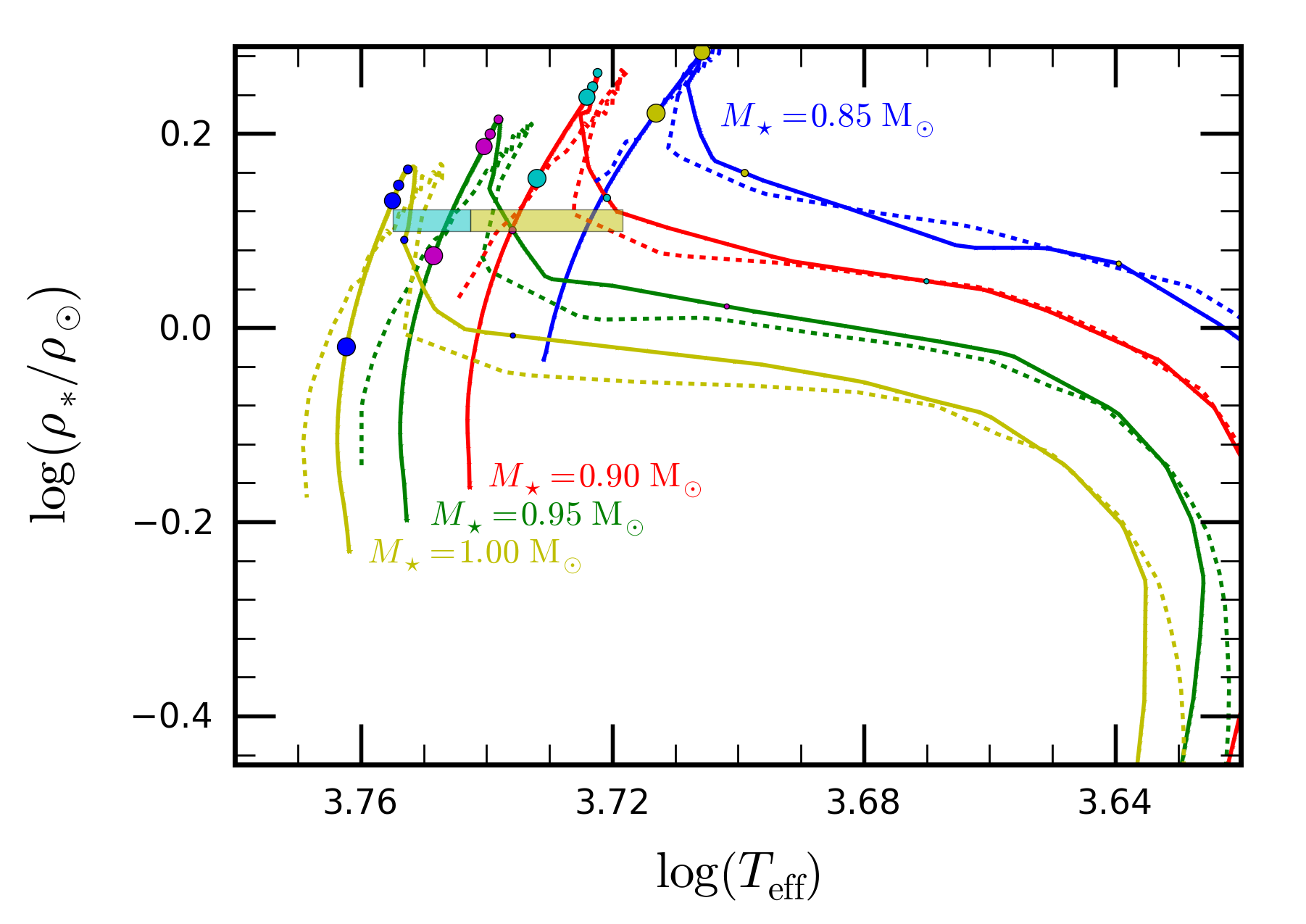}}}
\caption{Evolution tracks for stars of masses between 0.85 and 1.0$\rm
  M_{\odot}$ in the effective temperature vs. stellar density
  space. The $1\sigma$ observational constraints are shown as boxes:
  the box to the left (blue) corresponds to solutions that neglect the
  effect of spots. The two boxes (blue and yellow) correspond to the
  added uncertainty obtained when assuming that from 0 to 20\% of the
  stellar surface is covered by spots. CESAM tracks (plain) are
  compared to BCAH98 tracks (dashed). The models assume solar
  composition. Dots on the CEPAM tracks correspond to the following
  ages (small to large circles): 20, 32, 100, 500, 1000, 5000 Ma, respectively.}
\label{fig:evolution}
\end{figure}

The star's physical parameters ($M_\star$, $R_\star$, age) are
obtained by matching the constraints from Table~\ref{tab:obs} to a grid
of evolution models, as depicted in Fig.~\ref{fig:evolution}. The two
most important constraints of the problem are the star's effective
temperature and density. A third constraint (not shown on the plot) is
the star's {\em present} metallicity, which should be compared to the
one obtained from the evolution models that include diffusion. One
could include other constraints (such as that on $\log g$), but in the
present case, they are too weak to be useful.

The quality of the fit of any given model is given by its distance $n_{\sigma_\star}$
to the ellipsoid of constraints, measured in units of the standard error in the
constraints given by
\begin{equation}
n_{\sigma}=\left[\sum_i \left({X_{i}\over
      \sigma_{i}}\right)^2\right]^{1/2}, \nonumber
\end{equation}
where $X_i$ are the constraints and $\sigma_i$ their standard
deviation (assumed Gaussian).  The ellipsoid of constraints (of
dimension 2) is centered on ($\teff,\rho_\star$) and has semi-minor
and semi-major axes $k_{\rm 2D,n_{\sigma}} n_{\sigma}
(\sigma_{\teff},\sigma_{\rho_\star})$, respectively, where $k_{\rm
  2D,n_{\sigma}}$ is the quantile of a 2D gaussian law at the
equivalent level of confidence $n_{\sigma}$. In addition, $k_{\rm 2D, n_{\sigma}}
\sim 1.52,\, 2.49,\, 3.44$ for $n_{\sigma} = 1,2,3$
respectively. This normalization ensures that our solutions at 1,2,3
$\sigma$ have the correct probability of occurrence. However, for the
metallicity we adopt a relatively crude simplification: we consider as
valid only solutions for which the $\rm[Fe/H]$ value is within
1$\sigma_{\rm[Fe/H]}$ of the measured one. We tested that in the
particular case of CoRoT-2, considering yet larger errors in $\rm[Fe/H]$
has a negligible effect.  In the remainder of the paper, we present
models for which $n_{\sigma_\star}\le 1,2,3$, corresponding to
confidence levels of 68.3\%, 95.4\%, and 99.7\%, respectively.

Two possible values of the effective temperature are used: (i) in
the no-spot case, we assume that obtained by \citet{AmmlervonEiff+09} but with a slightly
larger error to account for possible systematic errors ($\teffs=5608\pm 80$\,K); (ii) when
including the effect of spots, we define a new temperature and its
associated error $\Delta_{\teff}$ to take into
account the presence of spots (from 0\% to 20\% of the area of the star)
\[\teffm =\teffsm {1+ (1-\chi_s)^{1/4}\over 2},\]
\[\Delta_{\teff}(n_{\sigma}) = \teffsm - \teffm + n_{\sigma} \sigma_{\teffs}.\]
We thus derive $\teff = 5224-5688$\,K at 1$\sigma$, $\teff=5144-5768$\,K at
2$\sigma$, and $\teff=5064-5848$\,K at 3$\sigma$. The errors are thus
intrinsically non-Gaussian. 
% lower that value by $0$ to
% $(1-0.2)^{1/4}$, i.e. $\teff=??\pm ??$\,K.

The constraints used for the stellar density and metallicity are those derived by \citet{Gillon+10}
and \citet{Alonso+08}, respectively (see Table~\ref{tab:obs}).

Figure~\ref{fig:evolution} shows evolution tracks in the
$(\log\teff,\log\rho_\star)$ plane for solar-composition stars of
between $0.85$ and $1.0\,\rm M_\odot$. For about $100$\,Ma, the stars contract
and heat up from a low-temperature, low-density initial state, the
so-called pre-main-sequence phase (PMS). They then reach a maximum in their
density and very gradually expand while on the main-sequence (MS). In the
case of CoRoT-2, the observational constraints on $\teff$ and
$\rho_\star$ are met either on the PMS, for ages $\sim 30$\,Ma, or at
much older ages in the MS phase. Only stars with masses between
roughly $0.84$ and $1.04\,\rm M_\odot$ intercept the box of constraints
at some point in their evolution. Figure~\ref{fig:evolution} also shows
that the presence of spots yields solutions at smaller masses than
when spots are not taken into account. The agreement between CESAM
and BCAH98 models is generally very good, with differences in
$\teff$ of generally $\sim 1\%$ or less, and differences in
$\rho_\star$ that can reach $\sim 10\%$ but in only a limited region of the
PMS evolution phase.

\begin{figure}
\centerline{\resizebox{8.5cm}{!}{\includegraphics{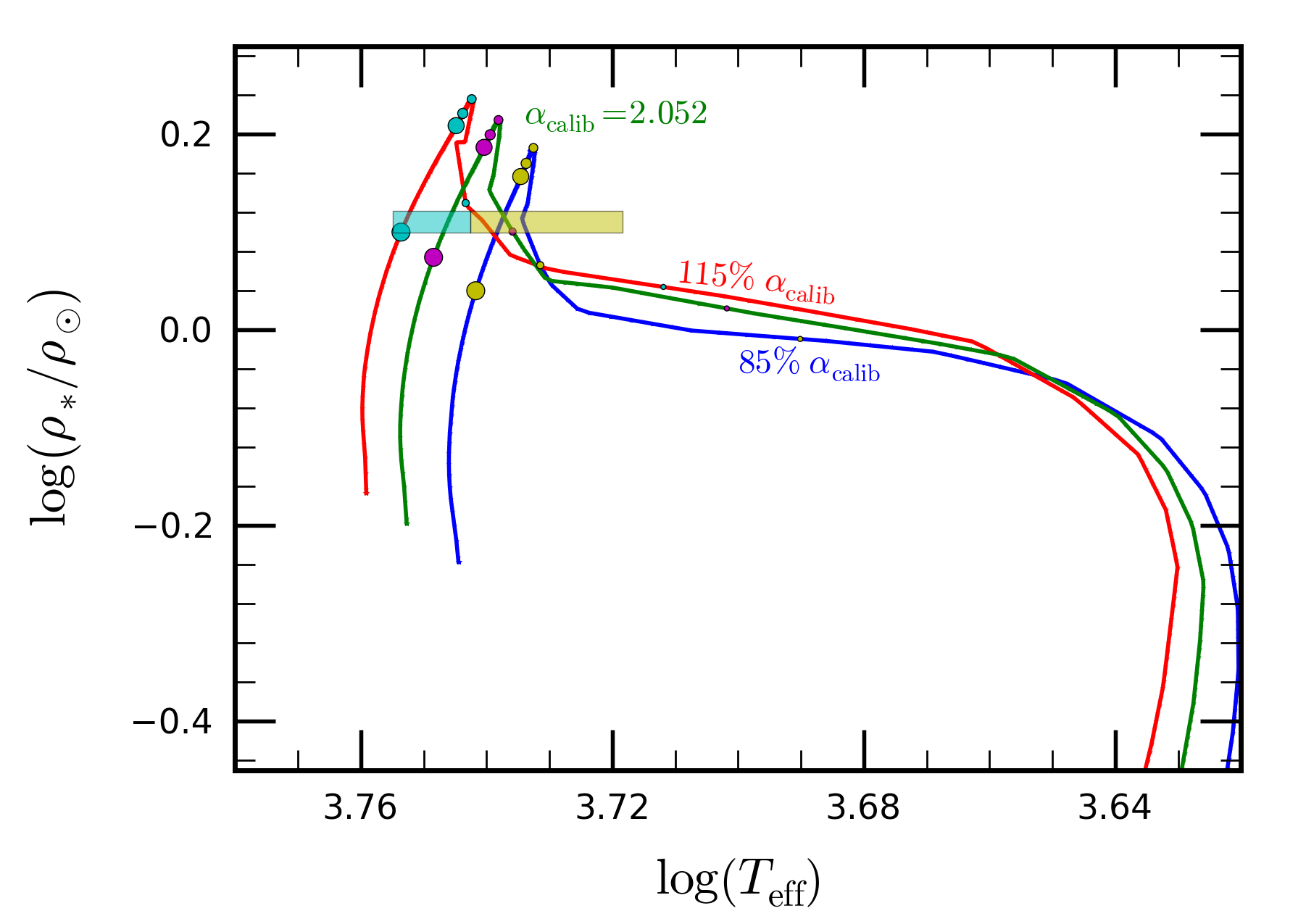}}}
\caption{CEPAM evolution tracks for a 0.95 $\rm M_{\odot}$ solar-composition
  star for three values of the mixing length parameter around the Sun
  calibrated value ($0.85\alpha_\odot$, $\alpha_\odot$,
  $1.15\alpha_\odot$) in the effective temperature vs. stellar density
  space. The symbols are as in Fig.~\ref{fig:evolution}.}
\label{fig:evolution_A}
\end{figure}

In Fig.~\ref{fig:evolution_A}, we explore the effect of a ``reasonable''
($\pm 15\%$) modification of the mixing length parameter $\alpha$ on
the evolution tracks. The effect is not negligible in terms of
its impact on both the effective temperature and the stellar
density. A higher value of $\alpha$ implies a more efficient energy
transport and therefore higher effective temperatures and generally a
faster evolution. Less intuitively perhaps, it leads to a higher
maximal stellar density at the early stages of the MS phase. However,
we emphasize that these models with modified mixing
lengths have not been calibrated and do not properly reproduce the
present Sun.

\begin{figure}
\centerline{\resizebox{8.5cm}{!}{\includegraphics{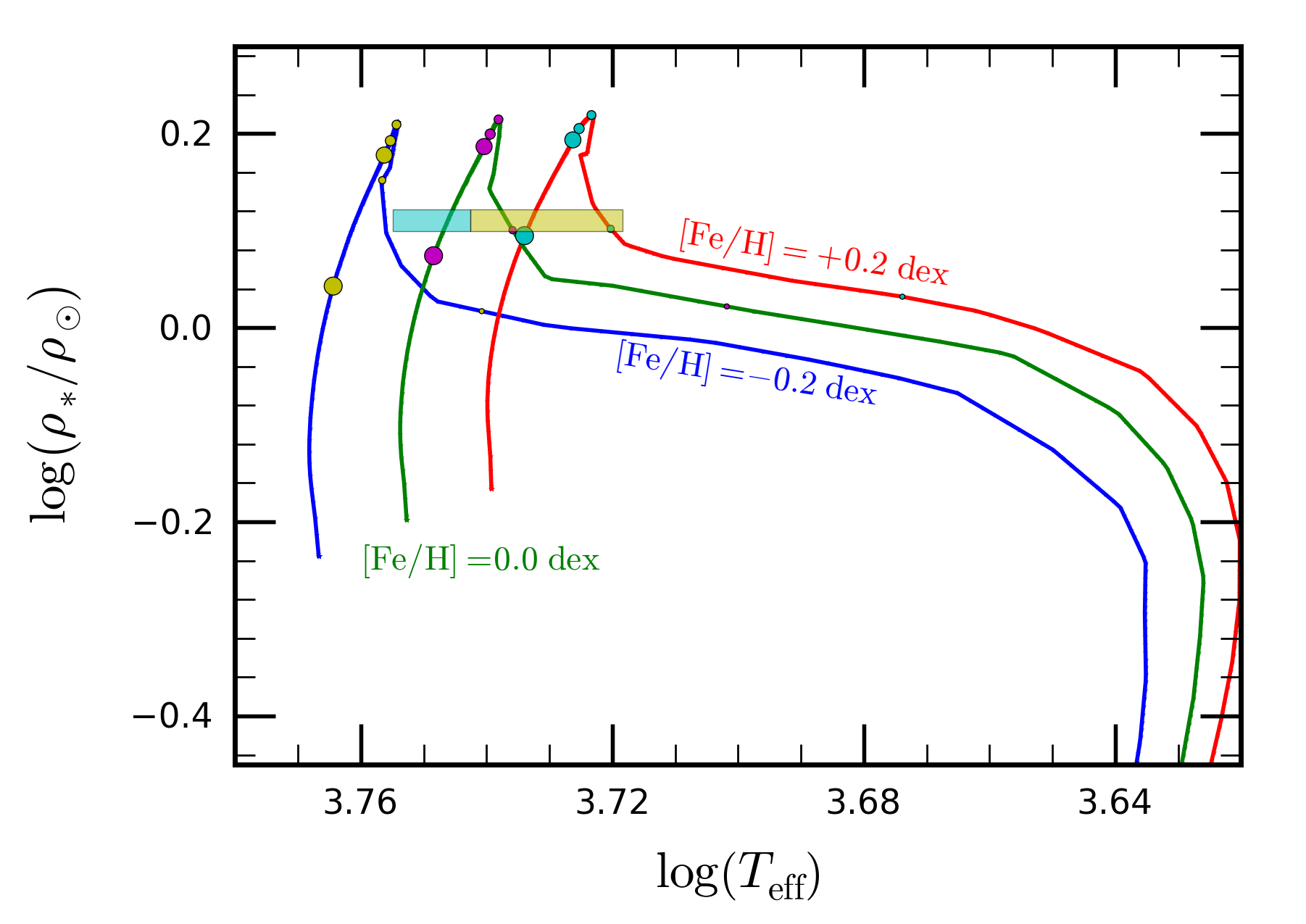}}}
\caption{CEPAM evolution tracks for a 0.95 $\rm M_{\odot}$
  star for three values of the metallicity ([Fe/H]=-0.2, 0, and 0.2,
  respectively) in the effective temperature vs. stellar density
  space. The mixing length parameter has been calibrated to the solar
  model. The symbols are as in Fig.~\ref{fig:evolution}.)}
\label{fig:evolution_Z}
\end{figure}

The consequences of metallicity variations are shown in
Fig.~\ref{fig:evolution_Z}. An increase in the [Fe/H] value by a factor
1/3 leads to a global decrease in the effective temperature by about
2\%, larger than the $1\sigma$ error in the measurements, but slightly
smaller than the uncertainty in $\teff$ obtained when including
spots. This effect is thus significant and shoud be included in the
search for solutions matching the observational constraints.

\begin{figure*}[htbp]
  \centerline{\resizebox{14cm}{!}{\includegraphics{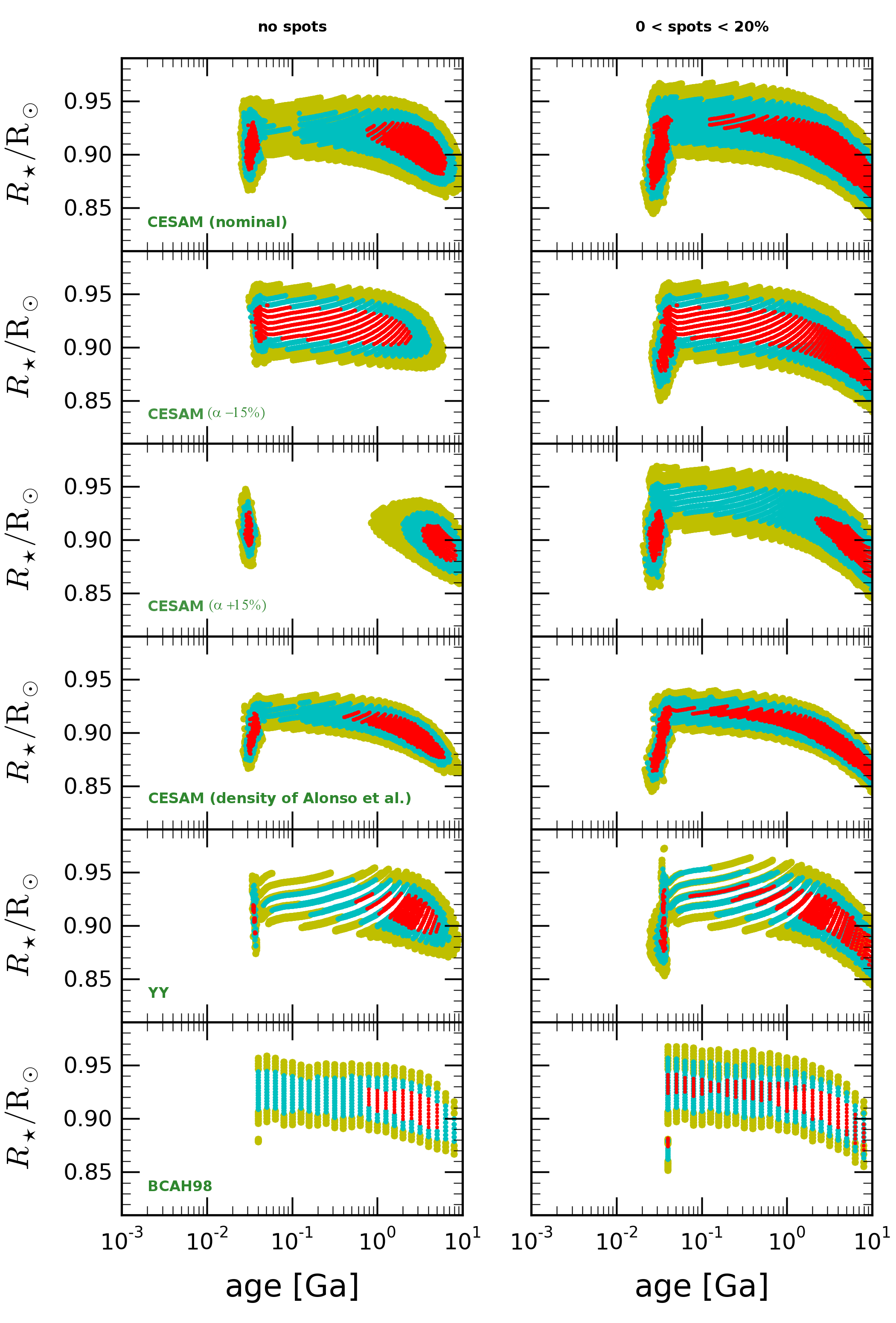}}}
  \caption{Constraints obtained for the age and radius of CoRoT-2
    with different assumptions. The left panels correspond to results
    obtained by neglecting the effect of spots. The right panels
    assume an additional uncertainty in the derived $\teff$ due to a 0\%
    to 20\% fraction of spots. From top to bottom, the panels are: a) Results
    obtained with the full CESAM calibrated evolution grid. b) Results
    obtained with CESAM with a mixing length parameter
    $\alpha=0.85\alpha_\odot$. c) Same as previously but with
    $\alpha=1.15\alpha_\odot$. d) Results obtained with the calibrated
    CESAM evolution grid but a constraint on the stellar density
    obtained from \citet{Alonso+08} instead of \citet{Gillon+10}. e)
    Results obtained with the YY tracks \citep{Demarque+04}. f)
    Results obtained with the BCAH98 tracks \citep{BCAH98}. The colored
    area corresponds to constraints derived from stellar evolution
    models matching the stellar density and effective temperature
    within a certain number of standard deviations: less than
    $1\sigma$ (red), $2\sigma$ (blue), or $3\sigma$ (yellow).}
\label{fig:rstar_age}
\end{figure*}

We present in Fig.~\ref{fig:rstar_age} the ensemble of solutions in the
$(R_\star ,{\rm age})$ space obtained with various assumptions. The
top panels correspond to our preferred solutions using our calibrated
CESAM evolution model and including all metallicities. At $3\sigma$, a
wide range of solutions is found that extends from ages between
$30$\,Ma and more than $10$\,Ga. Within $1\sigma$, two classes of
solutions are found, either on the PMS ($30-40$\,Ma) or on the MS (for
ages $>800$\,Ma with no spots, or $>100$\,Ma when including
spots). The range of stellar radii (which are directly proportional to
the planetary radii that are inferred) extend from $0.88$ to
$0.93\,\rm R_\odot$ at $1\sigma$, but the smallest values are obtained
for either the youngest or oldest solutions.

The other panels in Fig.~\ref{fig:rstar_age} highlight the consequences
of the different hypotheses on the solutions, when considering only solar
composition models. Varying the mixing length parameter has
consequences for the $1\sigma$ solutions: a lower $\alpha$ value
leads to a wider range of solutions at intermediate ages, while a higher $\alpha$
increases the separation between the very young and the very old
solutions. However, when considering the global $3\sigma$ envelope and
accounting for spots, the solutions are very similar.

The solutions obtained by using the $\rho_\star$ value of
\citet{Alonso+08} are quite similar to the nominal ones, but are
regarded as slightly over-constrained due to the assumption of a
circular orbit.

The last four panels in Fig.~\ref{fig:rstar_age} provide another test
of the robustness of the solutions by a comparison with YY and
BCAH98 evolution models. All models appear to be in excellent
agreement at least to the $2\sigma$ level. A minor difference is the
absence of $1\sigma$ PMS solutions in the no-spot case for BCAH98,
contrary to the CESAM and YY results.

Additional constraints on the stellar age may be derived from 
CoRoT-2 being a rapid rotator. According to \citet{MH08}, the 4.5
day rotation period with $\rm B-V=0.854$ \citep{Lanza+09} implies an
age typical of that of the Pleiades, i.e., $\sim 130\,$Ma. Given the
absence of stars with similar B-V and periods shorter than 8 days in
the Hyades ($\sim 625\,$Ma), we estimate that CoRoT-2 is less than
500\,Ma old, which thus restricts the ensemble of solutions from
Fig.~\ref{fig:rstar_age} quite significantly.

\begin{figure}[ht]
\centerline{\resizebox{8.5cm}{!}{\includegraphics{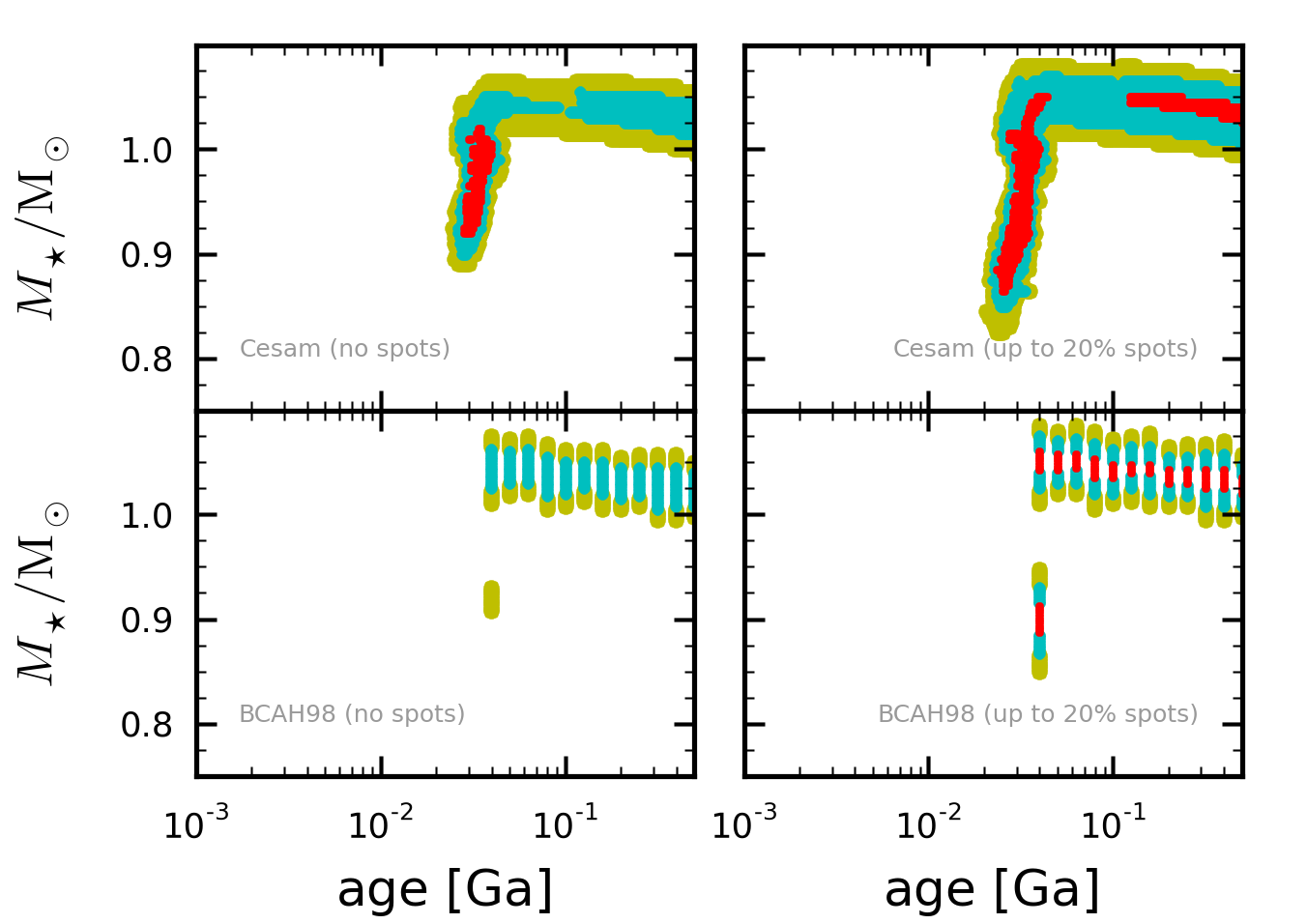}}}
\caption{Constraints obtained for the age as a function of mass of
  CoRoT-2. The left panels correspond to
  results obtained by neglecting the effect of spots. The right panels
  assume an additional uncertainty in the derived $\teff$ because of the presence of up to
  20\% of spots. The upper panels are calculated from CESAM evolution
  tracks. The lower panels are calculated from BCAH98 evolution
  tracks. Colors have the same meaning as in Fig.~\ref{fig:rstar_age}.}
\label{fig:ms}
\end{figure}

We now focus on the young ($<500$\,Ma) solutions and compare CESAM with
BCAH98 in the ($R_\star$, $M_\star$, age) parameter space. (We do not
show the comparisons with YY, since it closely ressembles CESAM). Figure~\ref{fig:ms} shows that in the no-spot case, the
$1\sigma$ solutions are limited to the PMS phase for CESAM, and there
are no solutions when using the BCAH98 models. At $2$ and $3\sigma$,
the solutions span the entire age range and both models yield very
similar results. We note that MS solutions
imply stellar masses slightly above that of the Sun, whereas PMS
solutions are distributed between $\sim 0.9$ and $1.0\rm\,M_\odot$.

\begin{figure}[ht]
\centerline{\resizebox{8.5cm}{!}{\includegraphics{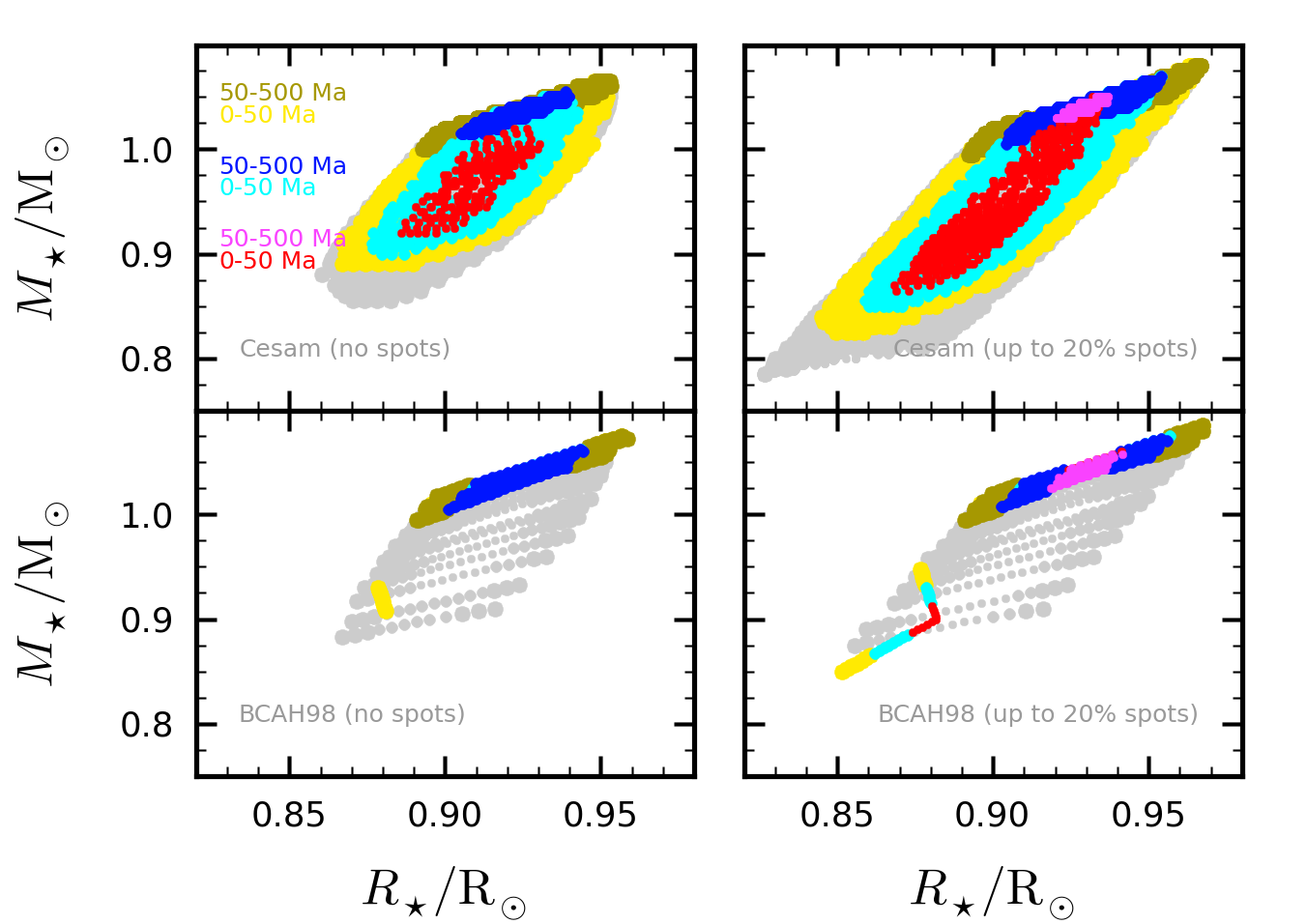}}}
\caption{Constraints obtained on the mass and radius of CoRoT-2. The
  panels and colors are as in Fig.~\ref{fig:ms}, except that the
  solutions are separated between those at young ($0-50$\,Ma) and old
  ($50-500$\,Ma) ages. Solutions obtained for ages above 500\,Ma are
  indicated in grey. }
\label{fig:msrs}
\end{figure}

Figure~\ref{fig:msrs} compares the solutions in the ($R_\star$,
$M_\star$) space. There is a clear positive correlation between the
two quantities. For ages older than 50\,Ma, the solutions are
confined to high mass values and there is a very good agreement
between CESAM and BCAH98. At young ages however, as seen in previous
diagrams, the CESAM and BCAH98 solutions differ. The influence of the
presence of spots is relatively small.

\begin{table*}[ht]
%\begin{center}
\caption{Derived mass and radius of the star CoRoT-2 for two different
  inferred ages.}
\label{tab:star}
\begin{tabular}{rcccc}
\multirow{2}*{Model} & \multicolumn{2}{c}{no spot} & \multicolumn{2}{c}{0-20\%
spots}\\
% \cline{2-5}
& $M_\star/M_\odot$ & $R_\star/R_\odot$ & $M_\star/M_\odot$
& $R_\star/R_\odot$\\
\hline
\hline
\multicolumn{5}{c}{\textit{0-50 Ma}}\\
% \hline
CESAM & $0.97 \pm 0.04$ & $0.91 \pm 0.02$ & $0.95 \pm 0.07$ & $0.90 \pm 0.03$\\
BCAH98 & $1.04 \pm 0.06\ (2\sigma)$ & $0.93  \pm 0.03\ (2\sigma)$ & $0.96 \pm
0.08$  & $0.90 \pm 0.03$\\
\multicolumn{5}{c}{}\\
\multicolumn{5}{c}{\textit{50-500 Ma}}\\
% \hline
CESAM & $1.03 \pm 0.01\ (2\sigma)$ & $0.92 \pm 0.01\ (2\sigma)$ & $1.04 \pm
0.01$ & $0.93 \pm 0.01$\\
BCAH98 & $1.03 \pm 0.02\ (2\sigma)$ & $0.92  \pm 0.02\ (2\sigma)$ & $1.04 \pm
0.02$  & $0.93 \pm 0.02$\\
\hline
%  \multirow{2}*{\rotatebox{90}{CESAM + BCAH98}}
%         & $M_\star/M_\odot$ & $ \pm 0.03$ & $ \pm 0.03$\\
%         & $R_\star/R_\odot$ & $  \pm 0.01$ & $ \pm 0.01$\\
% \hline
\end{tabular}
%\end{center}
\end{table*}

The results in terms of the stellar mass and radius are summarized in
Table~\ref{tab:star} for the different age classes. When no solutions
were found within $1\sigma$, the $2\sigma$ solutions are indicated.

\subsection{Constraints on the planet's physical parameters}

Physical parameters for the planet are derived from the solution for
the star using the {\modif orbital} period $P_{\rm orb}$, the radii ratio
$k=R_{\rm p}/R_\star$, and the semi-amplitude star velocity $K$ (see
Table~\ref{tab:obs}) using the following equations \citep[e.g.][]{Sozzetti2007, Beatty2007}
\begin{eqnarray}
 R_{\rm p} &=&  k R_\star,\\
 M_{\rm p} &=& \left(\frac{P_{\rm orb}}{2\pi G}\right)^{1/3} (M_\star+M_{\rm
p})^{2/3}
K \frac{\sqrt{1-e^2}}{\sin i} ,\label{eq:mp}
% M_p &=& M_\star^{2/3} K \left(\frac{P}{2 \pi G}\right)^{1/3}
\end{eqnarray}
where $G$ is the gravitational constant, $e$ the orbital eccentricity, and
$i$ the orbital inclination projected along the line of sight.
%The \autoref{eq:mp} is solved by multiple iterations.

\begin{table*}
%\begin{center}
\caption{Derived mass and radius of the planet CoRoT-2b for two different
  inferred ages.}
\label{tab:planet}
\begin{tabular}{rcccc}
\multirow{2}*{Model} & \multicolumn{2}{c}{no spot} & \multicolumn{2}{c}{0-20\%
spots}\\
% \cline{2-5}
& $M_{\rm p}/M_{\rm Jup}$ & $R_{\rm p}/R_{\rm Jup}$  & $M_{\rm p}/M_{\rm Jup}$
& $R_{\rm p}/R_{\rm Jup}$ \\
\hline
\hline
\multicolumn{5}{c}{\textit{0-50 Ma}}\\
% \hline
CESAM & $3.50 \pm 0.20$ & $1.52 \pm 0.04$ & $3.45 \pm 0.27$ & $1.50 \pm 0.06$\\
BCAH98 & $3.68 \pm 0.25\ (2\sigma)$ & $1.54 \pm 0.06\ (2\sigma)$ & $3.48 \pm
0.30$ & $1.50 \pm0.06$\\
\multicolumn{5}{c}{}\\
\multicolumn{5}{c}{\textit{50-500 Ma}}\\
% \hline
CESAM & $3.65 \pm 0.13\ (2\sigma)$ & $1.54 \pm 0.03\ (2\sigma)$ & $3.67 \pm
0.13$ & $1.55 \pm0.03$ \\
BCAH98 & $3.65 \pm 0.16\ (2\sigma)$ & $1.54 \pm 0.04\ (2\sigma)$ & $3.67 \pm
0.16$ & $1.55 \pm0.04$\\
\hline
%  \multirow{2}*{\rotatebox{90}{CESAM + BCAH98}}
%         & $M_\star/M_\odot$ & $ \pm 0.03$ & $ \pm 0.03$\\
%         & $R_\star/R_\odot$ & $  \pm 0.01$ & $ \pm 0.01$\\
% \hline
\end{tabular}
%\end{center}
\end{table*}

\begin{figure}
\centerline{\resizebox{8.5cm}{!}{\includegraphics{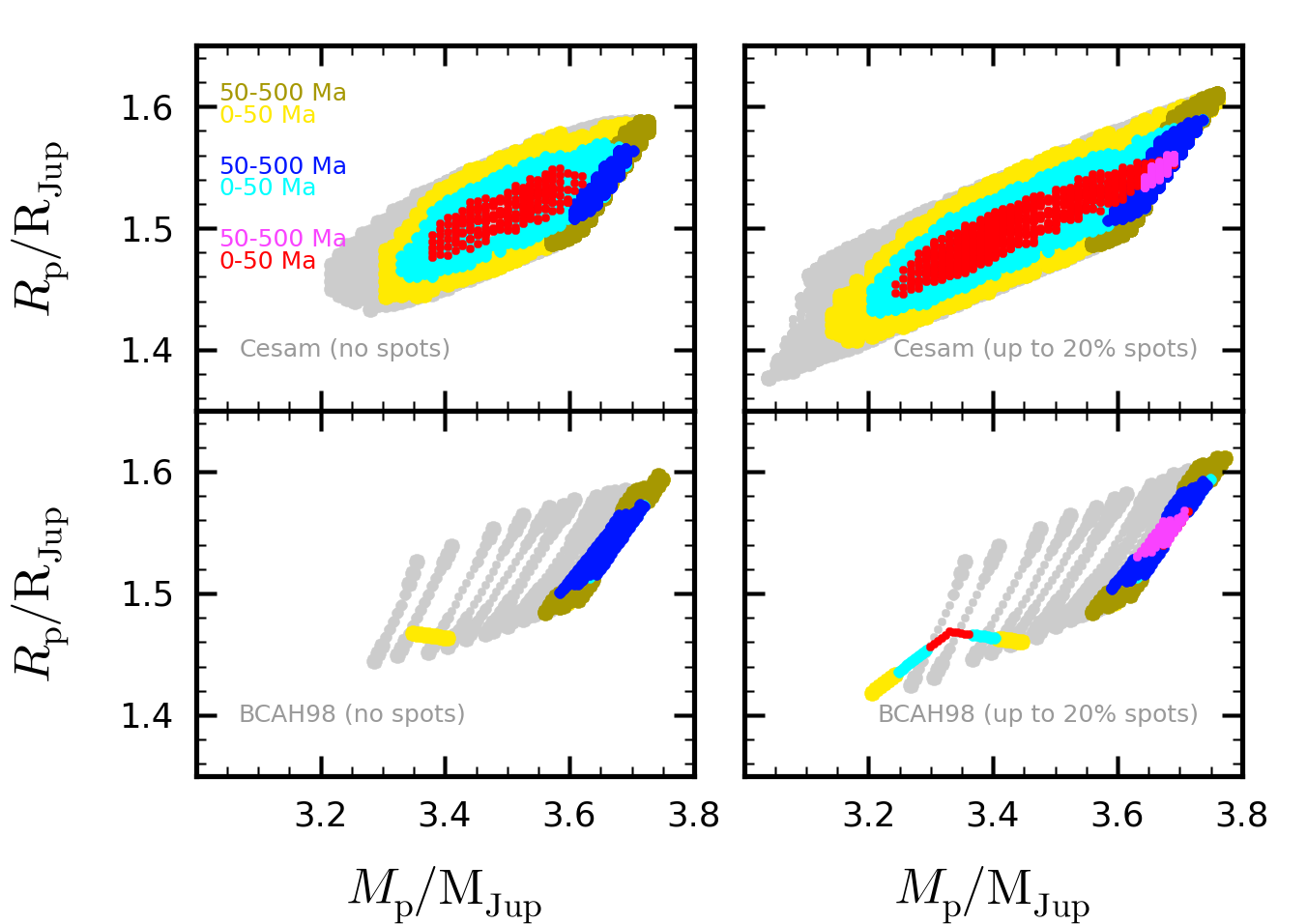}}}
\caption{Constraints obtained on the mass and radius of CoRoT-2b. The
  panels and colors are as in Fig.~\ref{fig:msrs}. $\rm R_{\rm
    Jup}\equiv 71492\,km$}
\label{fig:mprp}
\end{figure}

The results in terms of the inferred planet mass and radius are shown
in Fig.~\ref{fig:mprp} and listed in Table~\ref{tab:planet}. Our inferred
planet masses are slightly higher than obtained by \citet{Alonso+08}
but in good agreement with the \citet{Gillon+10} results. The
corresponding planetary sizes are larger than in both studies
because we account for the effect of spot occultations by
the transiting planet. The new classes of solutions for the star on
the PMS, at ages of between 30 and 40 Ma allow however for slightly smaller
$R_{\rm p}$ values than the standard MS solutions.

\section{Planetary evolution models \label{sec:planet_models}}

\subsection{Modeling procedure}

Our planet evolution models are calculated using CEPAM, a code originally derived from CESAM but accounting for the physics that is specific to planets \citep{GM95}. We adopt the same two hypotheses as models of other transiting exoplanets \citep[e.g.][]{Guillot08}, namely: (i) We use the equation of state for hydrogen and helium from \citet{SCvH95} and a slightly larger value of the helium mass-mixing ratio ($Y=0.30$) to account for the presence of heavy elements; and (ii) interior Rosseland opacities are calculated from \citet{Allard+01}. The outer boundary condition is slightly modified however to allow for the determination of the impact of atmospheric properties on the planetary evolution. Following \citet{Guillot10} \citep[see also][]{Hansen08}, we use a $T(\tau)$ model for the globally-averaged temperature field in the atmosphere
\begin{eqnarray}
\overline{T^4}&=&{3\tint^4\over 4}\left\{{2\over 3}+\tau\right\}+{3\teq^4\over 4}\left\{{2\over 3}+\right.\nonumber\\
&&\left. {2\over 3\gamma}\left[1+\left({\gamma\tau\over 2}-1\right)e^{-\gamma\tau}\right]+
{2\gamma\over 3}\left(1-{\tau^2\over 2}\right)E_2(\gamma\tau)\right\},
\label{eq:t4-global}
\end{eqnarray}
where $\tau$ is the optical depth at thermal wavelengths, $E_2$ is the exponential integral function of order 2, $4\pi R_{\rm p}^2\sigma \teq^4$ is the stellar luminosity received by the planet, $4\pi R_{\rm p}^2\sigma \tint^4$ is the planet's intrinsic luminosity and $\gamma\equiv\kapv/\kapth$ is the greenhouse factor equal to the ratio of the mean visible opacity to the mean thermal opacity.  In addition, we assume that $P=(\kapth/g)\tau$, and link atmospheric to interior models at the 10\,bar pressure level.

The values of the coefficients $\kapth$ and $\kapv$ are parameterized from detailed radiative transfer calculations, as described in \citet{Guillot10}. These, and hence the atmospheric models in general are very uncertain due to the weak constraints on the chemical composition of these atmospheres, the unknown cloud-coverage, the difficulty in estimating how atmospheric dynamics transports heat and chemical elements...etc. In any case, the values that most closely reproduce the models of \citet{FLMF08} in similar irradiation conditions are $\kapth=10^{-2}\rm\,g\,cm^{-2}$ and $\kapv=4\times 10^{-3}\rm\,g\,cm^{-2}$. These values yield evolution models that closely match our previous calculations \citep[e.g.][]{Guillot08}. However, to match the observational constraints from measured brightness temperatures we adopt as a baseline scenario with an increased thermal opacity $\kapth=1.5\times 10^{-2}\rm\,g\,cm^{-2}$ (see Sect.~\ref{sec:atm}).

The transit radius corresponding to the level for which the chord optical depth is equal to $2/3$ is calculated as in \citet{Guillot10} using our assumed visible opacity coefficient $\kapv$. Because of the large size of CoRoT-2b and its relatively significant mass, we choose to ignore the possibility that a central dense core is present and only consider a fully gaseous planet of approximately solar composition. An increase in the mean molecular weight and/or presence of a core would generally lead to a smaller planet size and thus increase the problem in reproducing the observations.

When considering models that include explicit tidal dissipation, we also solve the combined dynamical and structural evolution of the star/planet system including tides as described in the Appendix. We follow changes in the structural planetary parameters (radius $R$, pressure $P$, temperature $T$, intrinsic luminosity $L$) as well as in the dynamical parameters of the system (semi-major axis $a$, orbital eccentricity $e$, stellar spin $\Omega_1$, planetary spin $\Omega_2$).  At each evolution timestep, a new value of the tidal heating rate $H_2$ is calculated by solving implicitly the equations for the dynamical evolution of the system. This heating rate is used for the subsequent structural calculations, and assumed to be dissipated at the center of the planet, so that $L(r=0)=H_2$. (For a discussion of how the depth of the dissipation affects the planet's structure and evolution, see \citet{GS02}.) The orbital evolution also modifies the atmospheric boundary condition by altering the irradiation flux. The equilibrium temperature $\teq$ is hence recalculated at each timestep.  Our approach to the tidal dissipation calculation is thus similar to that chosen in other calculations \citep{IB09, MFJ09}, but it is based on the equations derived by \citet{BO09} instead of those of \citet{JGB08a}. The main differences are that the relations are valid for higher values of the eccentricity \citep[see also][]{LCBL10} and include the secular evolution of stellar and planetary spins.

\subsection{Standard evolution models}

\begin{figure}
\centerline{\resizebox{8.5cm}{!}{\includegraphics{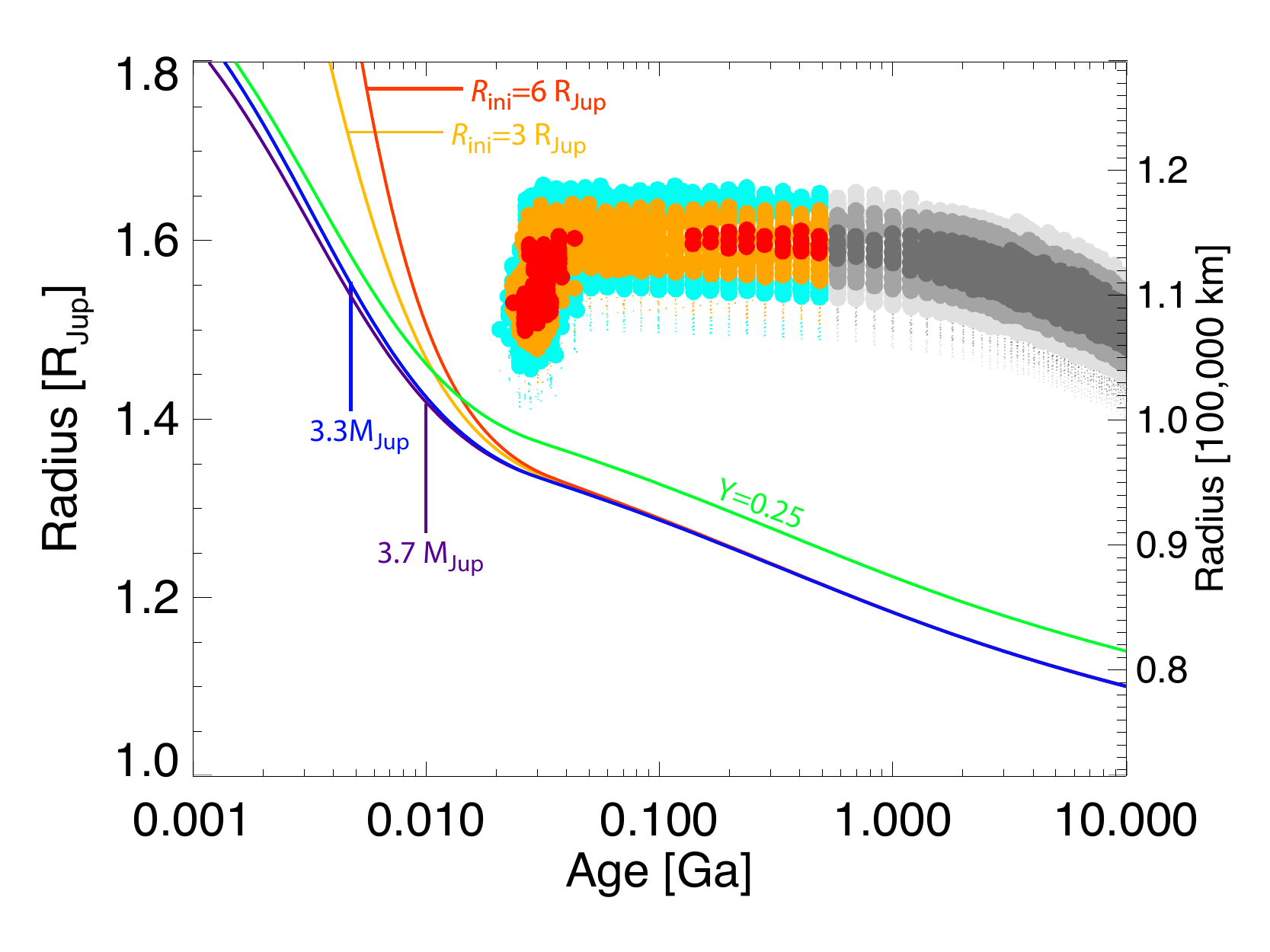}}}
\caption{Contraction of CoRoT-2b compared to its measured radius
  and inferred age. Our fiducial model has an initial radius $R_{\rm
    ini}=2\,$\rjup, mass $M_{\rm p}=3.5\,$\mjup, and equivalent helium
  mass-mixing ratio $Y=0.30$. The evolution for models with different
  initial radii $R_{\rm ini}$ (3 to 6\,\rjup), different masses ($3.3$
  to $3.7\,$\mjup), or a different value of $Y$ ($0.25$) are shown as labeled.}
\label{fig:evol-corot2}
\end{figure}

We compare in Fig.~\ref{fig:evol-corot2} observational constraints on age and planetary size to standard evolution models with slightly different assumptions about mass, initial planetary radius, and helium content. The standard models are defined by the planet only contracting from an initial radius $R_{\rm ini}$ as a result of the loss of its internal entropy: the only reservoir of energy is the initial gravitational energy  $\int Gm/r\ dm$. The atmospheric boundary condition corresponds to our baseline case.

Our fiducial model has a mass of $M=3.5$\,\mjup, an initial radius
$R_{\rm ini}=2$\,\rjup, and an equivalent helium mass mixing ratio
$Y=0.30$. It falls short of reproducing the inferred radius by $\sim
20$\% or more, except at young ages: for pre-main-sequence solutions at $30-40$\,Ma, the discrepancy is reduced to about 10\%. Within the error bars, models with different masses have almost identical evolutions and therefore planetary mass is not a significant parameter in the problem (this is because CoRoT-2b lies in the particular regime of mass for which the radius is almost independent of mass, which for isolated objects corresponds to a maximum in the mass-radius relation -see \citet{Guillot05}). In a similar way, changing the initial radius affects the evolution only in the first million years: the initial conditions are rapidly forgotten. One may wonder whether a different composition, in particular a different helium abundance, would have a stronger effect on evolution? As shown in Fig.~\ref{fig:evol-corot2}, models with $Y=0.25$ indeed yield a $\sim 4\%$ larger size than for $Y=0.30$, but this again falls short in explaining the large size of the planet.

These calculations confirm that CoRoT-2b is an anomalously large
planet, a result already obtained by \citet{Alonso+08},
\citet{Leconte+09}, and \citet{MFJ09}. However it also demonstrates
the fact that the planet's young age is likely to be a crucial factor
in explaining its size, both because of the possibility that the planet
is initially quite large, and because our stellar
evolution models yield solutions at 30-40\,Ma that are closer to the theoretical evolution tracks than at any later times.

\subsection{CoRoT-2b among its peers}

\begin{figure}
\centerline{\resizebox{8.5cm}{!}{\includegraphics{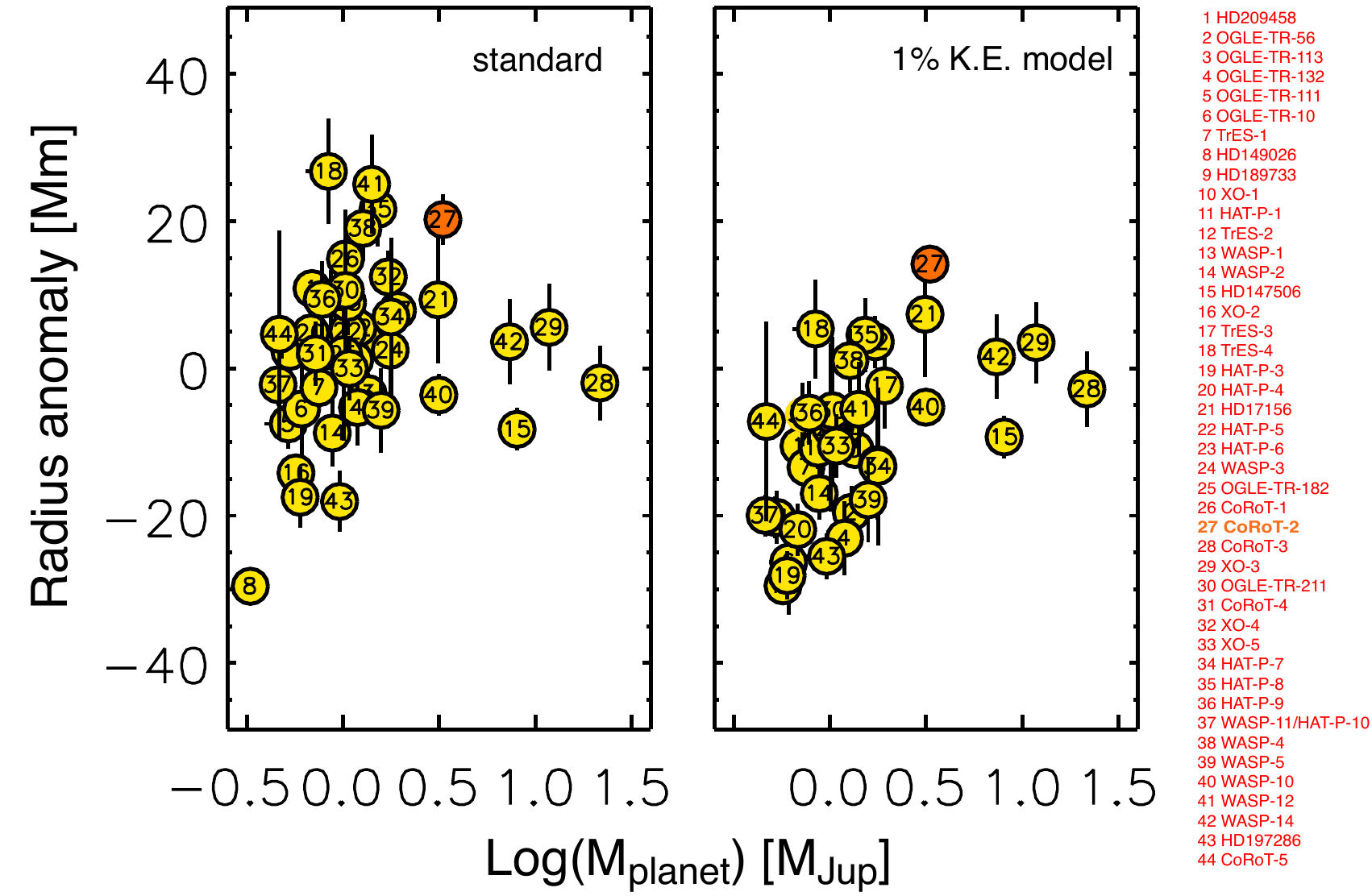}}}
\caption{Radius anomaly (difference between observed and modeled
  radius --see text--) as a function of planetary mass (in log) for a
  selection of known transiting giant planets. {\it
    Left panel:} Standard models (including stellar irradiation but no
  extra heat source) are used for the comparison. {\it Right panel:}
  Models assume an extra heat source at the planet's center equivalent
  to 1\% of the incoming stellar heat flux (see Guillot \& Showman
  2002). CoRoT-2b is labeled ``27''. It is the most anomalously
  large planet on the right panel.}
\label{fig:Ranomaly}
\end{figure}

It is instructive to compare CoRoT-2b to an ensemble of other transiting giant planets. Among these, CoRoT-2b may not be the largest (it is smaller than CoRoT-1b, HAT-P-8b, TrES-4b, and the present record-holder WASP-12b), but it remains the most difficult to reconcile with present-day models. This is most easily shown by ranking the planets in terms of their radius anomaly, i.e. the difference between its measured radius and the one predicted by models of a pure solar-composition planet of the same mass and age \citep{Guillot+06}.  As shown by Fig.~\ref{fig:Ranomaly}, when using standard models, CoRoT-2b has a positive, large radius anomaly of 20000\,km, but still smaller than that of HAT-P-8b, TrES-4b, and WASP-12b. However, for these last three planets, their large radius can be explained within the error bars by an additional heat source equivalent to 1\% of the incoming stellar luminosity deposited at the planet's center (see Guillot 2005 and references therein). As shown in the right panel of Fig.~\ref{fig:Ranomaly}, this is not true for CoRoT-2b: because of its large mass, the planet tends to contract rapidly and therefore requires special conditions to explain its large size.

\subsection{CoRoT-2b's atmosphere}\label{sec:atm}

We now consider how models of the planet's atmosphere affect its evolution. Remarkably, secondary transits of CoRoT-2b were detected in the optical from CoRoT lightcurves \citep{Alonso+09, Snellen+10} and in the infrared from Spitzer IRAC observations \citep{Gillon+10} as well as ground-based measurements \citep{ADKR10}. These directly probe the planetary atmosphere and are thus key constraints of the outer boundary conditions used in the evolution modeling.

The fluxes inferred from these measurements are equivalent to brightness temperatures of $1325\pm 180$\,K at $8\,\mu $m and $1805\pm 70$\,K at $4.5\,\mu $m \citep{Gillon+10, Snellen+10}. Additional ground-based measurements yield $T_{\rm b}=1890^{+260}_{-350}\,$K in the K$_s$ band ($\sim 2.1\,\mu$m) and an upper limit of $2250\,$K in the H band ($\sim 1.6\,\mu$m) \citep{ADKR10}. In the optical, independent studies yield brightness temperatures that are very similar within error bars, i.e. $2120^{+90}_{-110}\,$K \citep{Alonso+09,Alonso+10} and $2170\pm{50}\,$K \citep{Snellen+10}. A crucial consequence that can be derived is that {\em independently of hypotheses about the atmospheric composition
  and opacity sources, the day-side atmosphere of CoRoT-2b is
  characterized by physical temperatures that are as low as $1325\pm
  180\,$K, and at least as high as $1805\pm 70\,$K.} The case of the optical brightness measurements are more complex to interpret because they may arise from either direct emission or a reflection of the incoming stellar flux \citep{Alonso+09,Snellen+10}. In the limit of a geometric albedo of $0.2$, the flux would be entirely due to direct reflection and thus provide no information about the atmospheric temperature profile. In contrast, a zero albedo would imply that the flux in the optical is thermal emission from the planet, and that the corresponding temperatures are high.

\begin{figure}
  \centerline{\resizebox{8.5cm}{!}{\includegraphics{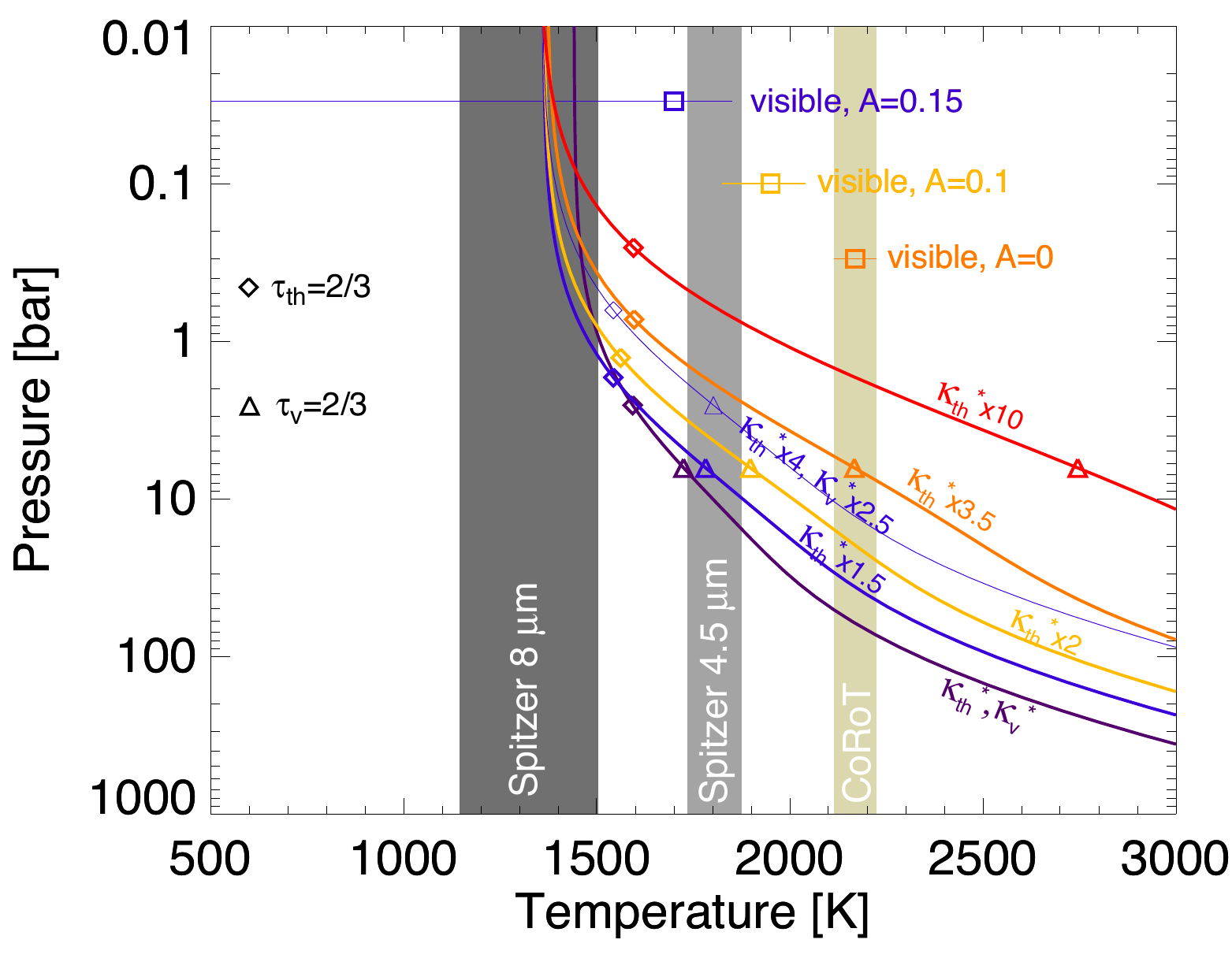}}}
\caption{Possible atmospheric pressure-temperature profiles for
  CoRoT-2b compared to observational constraints. The brightness
  temperatures obtained by Spitzer IRAC at 4.5 and 8$\,\mu$m
  \citep{Gillon+10} and by CoRoT in the optical
  \citep{Alonso+10,Snellen+10} are indicated by vertical grey
  bands. In the case of the optical brightness, the corresponding
  physical atmospheric temperatures depend on the atmospheric
  geometric albedo ($A=0$ to $0.15$), and are indicated by squares and
  with error bars. Temperature profiles are calculated on the basis of a
  semi-grey analytical model \citep{Guillot10}, with fiducial values
  of the thermal and visible opacities $\kapth^*=0.01\,\rm
  cm^2\,g^{-1}$, $\kapv^*=0.004\,\rm cm^2\,g^{-1}$, and $\tint=1000$\,K. The different
  lines correspond to different values of these coefficients (as labeled). The values of the assumed Bond albedo are $A=0$ (purple, orange, red curves), $A=0.1$ (yellow curve) and $A=0.15$ (blue curves). The
  levels for which thermal and visible optical depths
  equal 2/3 are indicated by diamonds and triangles, respectively.}
\label{fig:atm-corot2b}
\end{figure}

These temperature constraints are compared in Fig.~\ref{fig:atm-corot2b} to temperature profiles calculated for CoRoT-2b in the framework of our semi-gray analytical model assuming a full redistribution of the incoming stellar flux \citep[see Eq.~(\ref{eq:t4-global}) and ][]{Guillot10}. The value of the intrinsic flux $\tint=1000\,$K was chosen to match that of models with a size $\sim 1.5\,R_{\rm Jup}$, as observed. Using values of the thermal and visible opacities calibrated to detailed atmospheric models \citep{FLMF08} (labeled $\kapth^*$, $\kapv^*$ in the figure), we derive a temperature profile that is difficult to reconcile with the Spitzer and CoRoT brightness temperatures: the temperature range spanned by the atmosphere at mean optical levels smaller than unity is small: from about 1400 to 1600\,K. This implies that to explain the observations with this model, one needs to invoke (i) an albedo of $\sim 0.2$, (ii) a high-opacity at $8\,\mu$m, and (iii) a very-low opacity (lower than at visible wavelengths) at $4.5\,\mu$m. This combination of factors appears to be unlikely. Furthermore, one should consider that the low 8$\,\mu$m temperature has to be emitted from high levels in the atmosphere. Because the radiative timescale is shorter at high altitudes \citep{SG02, IBG05}, its temperature may have to be closer to the dayside than global average, which would make the problem even worse.

For these reasons, we consider alternative models. We restrict ourselves to
models without a temperature inversion, both because this is not predicted
by dedicated radiative transfer models \citep{Gillon+10, Snellen+10},
and it would make the planetary radius problem more
severe. Figure~\ref{fig:atm-corot2b} presents a variety of alternative
profiles. The temperature range in the low-optical thickness part of
the atmosphere is directly related to the factor
$\gamma\equiv\kapv/\kapth$, with a low $\Gamma$ value implying a
strong greenhouse effect. We find that models that can more readily reproduce 
both the infrared and visible brightness temperatures
are those with $\gamma\approx 0.2$ to $0.4$. These models are also
consistent with the secondary transit of CoRoT-2b being
detected in CoRoT's red channel and not in the blue and green channels
\citep{Snellen+10}, implying that it indeed originates mostly from
direct planetary emission rather than stellar light reflection. They
are also consistent with their smaller implied albedos, in line with
the low limit $A_{\rm g}=0.038\pm 0.045$ obtained for HD209458b
\citep{Rowe+08}, a planet with a similar equilibrium temperature.
Larger greenhouse effects (i.e. the red curve in
fig.~\ref{fig:atm-corot2b}) are not consistent with the models because they would tend to
yield brightness temperatures that are larger than inferred.

Compared to traditional atmospheric models, this larger greenhouse effect may be achieved by several means: the presence of clouds, of photochemical products, or generally of minor species that are efficient infrared absorbers but are transparent at visible wavelengths. Complications arising from chromatic and dynamical effects should of course be taken into account, but are should not change qualitatively the conclusions of this work.

\subsection{Impact of atmospheric models on the planetary evolution}

\begin{figure}
\centerline{\resizebox{8.5cm}{!}{\includegraphics{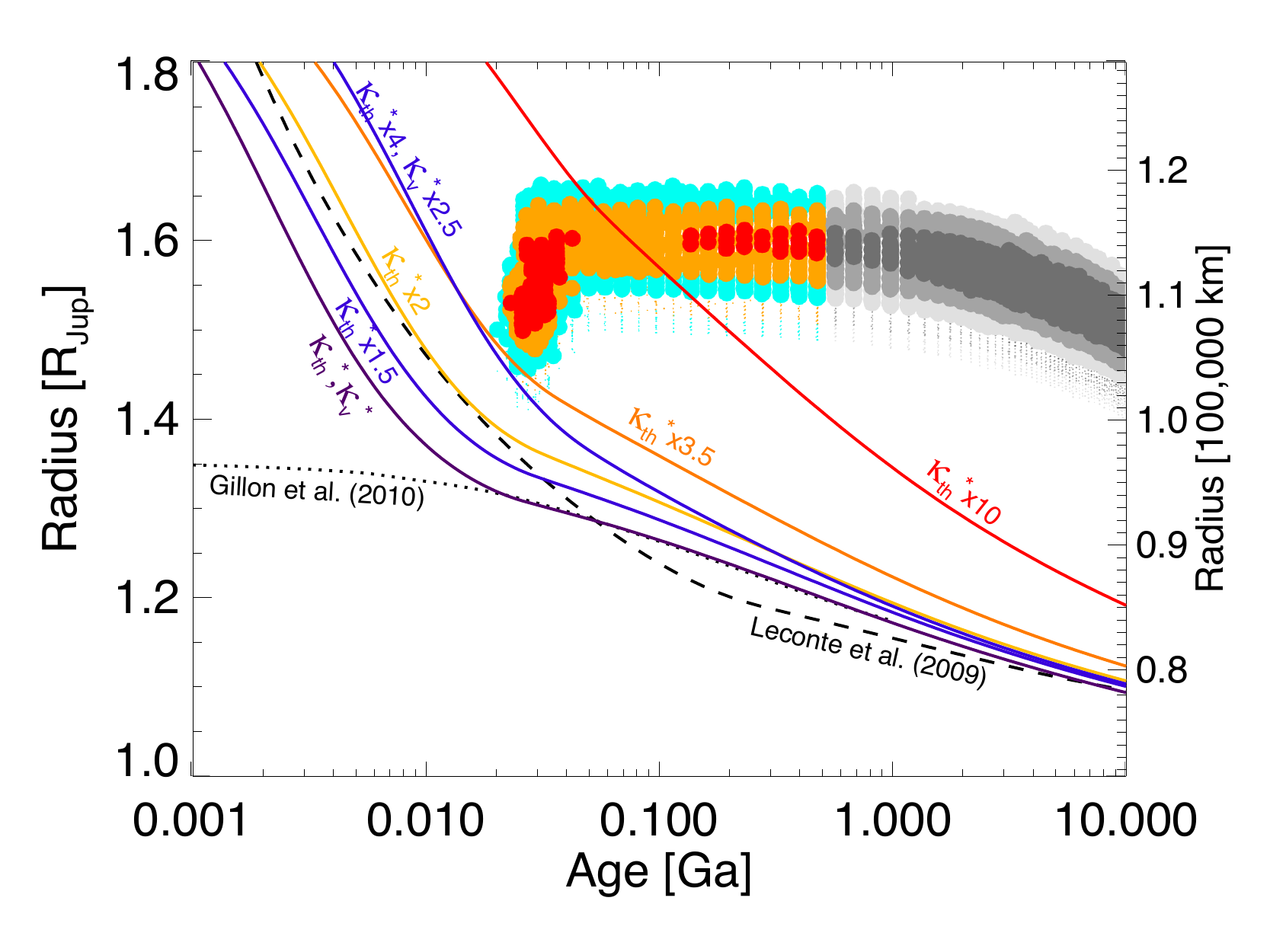}}}
\caption{Transit radius of CoRoT-2b as a function of age. The colored areas indicate constraints obtained from photometric, spectroscopic, radial-velocity data and stellar evolution models. The colors are a function of the distance to the effective temperature-stellar density constraint ellipse in standard error units: within $1\sigma$ (red), $2\sigma$ (yellow) and $3\sigma$ (blue). The star's activity implies an age smaller than $\sim 0.5\,$Ga, which indicated by solutions in shades of grey. The main constraint is obtained for a lightcurve analysis that includes the presence of spots. An analysis that does not account for spots yields a $\sim 3\%$ smaller radius, indicated by a dotted area (see text). The plain curves are obtained from standard evolution models using atmospheric boundary conditions that are parameterized by mean infrared $\kapth$ and visible $\kapv$ opacity coefficients, as in Fig.~\ref{fig:atm-corot2b}. The dashed and dotted curves are models from \citet{Leconte+09} and \citet{Gillon+10}, respectively. }
\label{fig:evol_compare_atms}
\end{figure}

Planetary evolution models calculated with different atmospheric boundary conditions are compared in Fig.~\ref{fig:evol_compare_atms} to standard evolution models for this planet available in the literature and to our inferred age/size constraints for CoRoT-2b. We first note that our baseline models (with opacities $\kapth^*$ and $\kapv^*$) are a close match to the models of \citet{Gillon+10} after about 20\,Ma, and that they reproduce relatively well the models of \citet{Leconte+09}, within 5\%, over the whole age range. In this last case, the differences in behavior may be attributed to our fiducial model having slightly lower infrared opacities and a smaller $\gamma$ value. Because the internal structure part of the calculation should be very similar, this highlights that even with similar hypotheses (a solar composition cloud-free atmosphere), detailed radiative transfer atmospheric models yield predictions that differ in a relatively significant way \citep[see also][]{Guillot10}.

As previously, most models are unable to reproduce the
constraints. However, two of them do intersect the constraint area at
young ages between $30$ and $40$\,Ma. These are models for which the thermal
atmospheric opacity has been increased by a factor $3.5$ to $4$,
relatively independently of visible opacities. The consequence of the
larger $\kapth$ is to reduce the amount of initial heat lost by the
planet at young ages. The intersection is small, but could become
larger if we considered that non-grey models may enable
the deep atmospheric temperatures to increase while the
visible brightness temperature remains constant, and also 
that the planet probably formed at least a few million years after the
star. As shown in Fig.~\ref{fig:evol_compare_atms}, for an even stronger
greenhouse effect (a smaller $\gamma$), the present size of the planet
may be reached at even older ages. However, this extreme model is unlikely 
because it disagrees with the inferred visible brightness temperature
(see \S~\ref{sec:atm}).

The present size of CoRoT-2b can thus be explained by the combination of a young age of between $30$ and $40$\,Ma, and additional opacity sources (gases/clouds) in the atmosphere. We note that \citet{BHBH07} also proposed an increase in atmospheric opacities to explain the large sizes of exoplanets. Our solution is similar, but we emphasize that this increase should concern more particularly the opacities at infrared wavelengths.

\subsection{Alternative recipes}

\begin{figure}
\centerline{\resizebox{8.5cm}{!}{\includegraphics{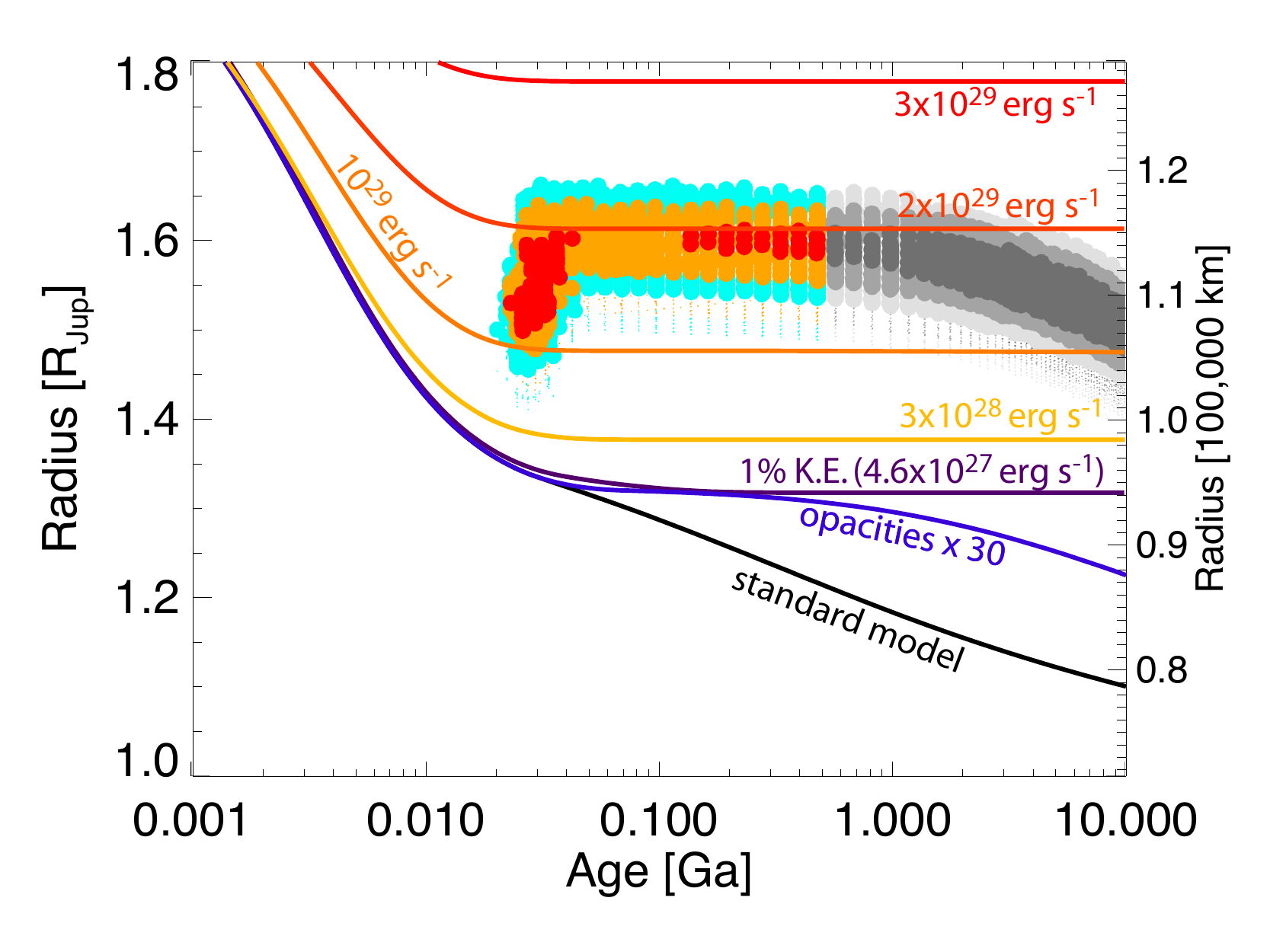}}}
\caption{Contraction of CoRoT-2b relative to its measured radius
  and inferred age (see fig.~\ref{fig:evol_compare_atms}). The evolution models are labeled as follows: {\it
    standard (black)}: irradiated, solar-composition planet with no extra heat
  source; {\it 1\% K.E. (blue)}: 1\% of the incoming stellar flux is assumed
  to be dissipated at the planet's center; {\it opacities$\times30$ (light blue)}:
  opacities have been artificially multiplied by 30 compared to the
  standard model; {\it $10^{29}$ (red)}: model in which $10^{29}\,\rm
  erg\,s^{-1}$ is deposited at the planet's center.}
\label{fig:compare_ctediff}
\end{figure}

Alternative models invoked to explain the large sizes of other exoplanets are unlikely to work. As shown in Fig.~\ref{fig:compare_ctediff}, the kinetic-energy dissipation model proposed by \citep{GS02} and used successfully for most known transiting planets \citep{Guillot+06,Guillot08} fails for CoRoT-2. This is also the case of a considerable -30 fold- increase in interior opacities that would also explain the sizes of most transiting planets \citep{Guillot08}. In fact, as already noted \citep{Alonso+08,Gillon+10}, the energy dissipation deep inside the planet required to explain the present-day radius is enormous, on the order of $10^{29}\,\rm erg\,s^{-1}$. This is about 30000 times the present intrinsic luminosity of Jupiter. It is also about 1/4th of the power that the planet receives from its parent star.

After that provided by stellar radiation, the most important potential source of energy is that taken from the planetary orbit. When moving CoRoT-2b from infinity to its present orbit, $\Delta E=GM_* M_{\rm p}/2a \approx 10^{45}\,$erg have to be dissipated. If this energy dissipation were to occur entirely in the planet, the maximum amount of time one would be able to maintain a $10^{29}\,\rm erg\,s^{-1}$ dissipation rate is $\sim 300$\,Ma.

\subsection{The effect of tides}

As originally proposed by \citet{BLM01} and \citet{GLB03} and later studied by many authors \citep[e.g.][]{JGB08a,IB09,MFJ09}, stellar tides provide a way to transfer gravitational energy from the planetary orbit into the planet and either slow its contraction, or even produce a size inflation. Models coupling the equations governing the dynamical evolution of the star+planet system with the physical planetary evolution rely however on a crucial assumption: that dissipation occurs at a sufficient depth in the planet interior, i.e. roughly within the planet's convective zone, deeper than a few 100 bars or so \citep[see][for a discussion]{GS02}. The mechanisms responsible for the dissipation are yet unknown, and may occur either high up in the atmosphere \citep{LTL97} or throughout the planetary interior \citep{OL04}.

Following \citet{Gillon+10}, we present models of the dynamical and physical evolution of the CoRoT-2 system caused by the action of stellar and planetary tides. We maximize the efficiency of the heat dissipation by assuming that it is entirely deposited at the center of the planet. We use the dynamical evolution equations derived by \citet{BO09} and include high order terms in eccentricity and equations for the evolution of the stellar and planetary spin (see Appendix). On the basis of the calculations by \citet{JGB08a}, we explore values of the tidal factor $Q_{\rm p}$ between $10^{5}$ and $10^6$, and of $Q_*$ of $10^5$ and higher.

\begin{figure}
\centerline{\resizebox{8.5cm}{!}{\includegraphics{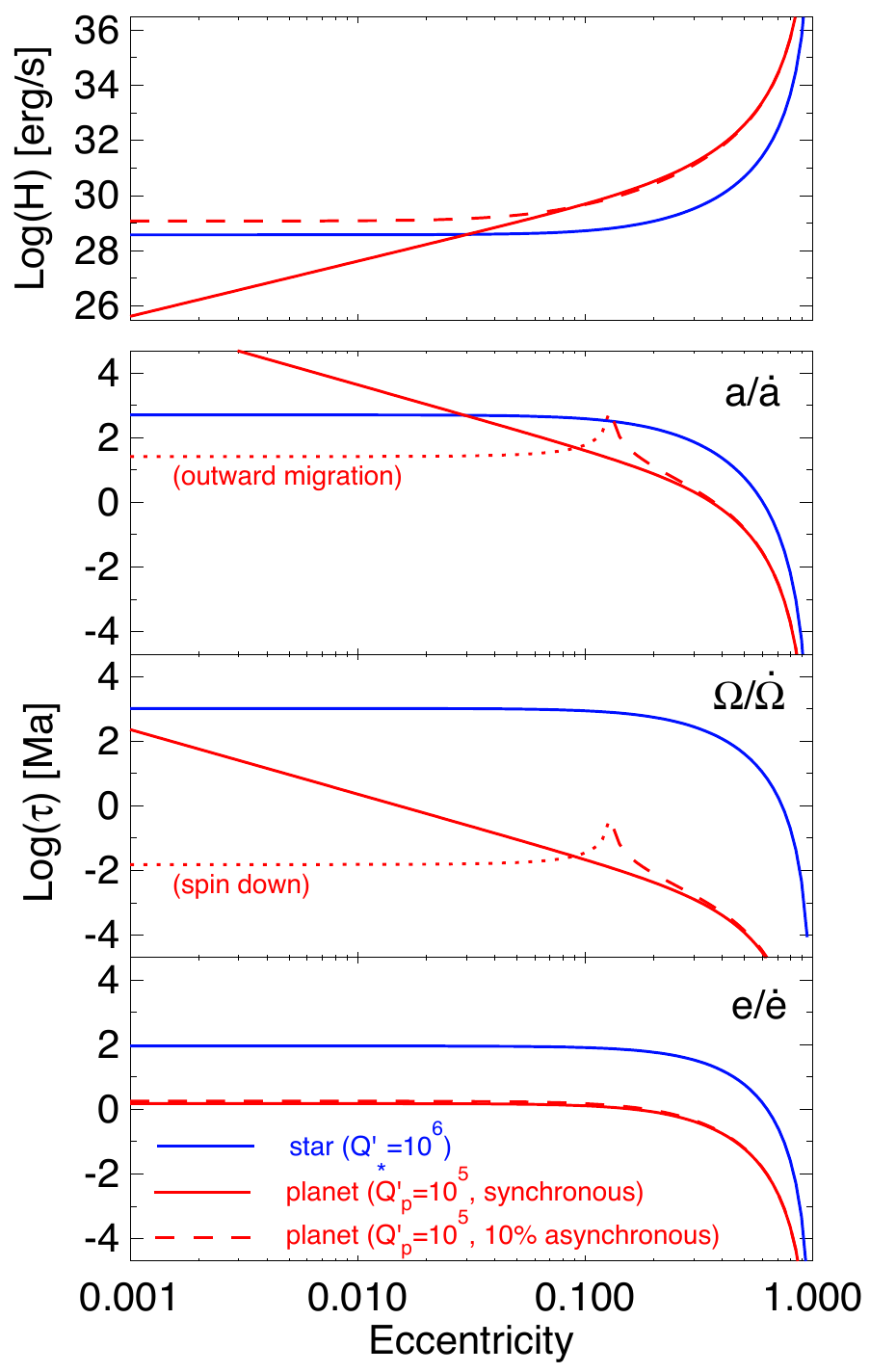}}}
\caption{Heating rates and dynamical timescales for the orbital evolution of the CoRoT-2 system as a function of the orbital eccentricity. The star and planet are assumed to have their present mass and orbital period, and tidal dissipation factors $Q'_*=10^6$, $Q'_{\rm p}=10^5$. The blue curves refer to the star, the red curves to the planet. The planet is assumed to be either synchronously rotating (spin period=orbital period) (plain lines), or to rotate 10\% faster than synchronous rotation (dashed lines). From top to bottom, the panels show various important quantities: (a) {\em Heating rates} due to tidal dissipation in the planet and in the star. (b) {\em Migration timescale} due to the star and the planet. Migration is inward except in the asynchronous case due to the planet's spin down for eccentricities smaller than $\sim 0.12$. (c) {\em Spin-up timescale}: The star is spun up by the planet in all cases. The planet is generally spun up, except in the asynchronous case (see text). (d) {\em Circularisation timescale}: The evolution towards a circular orbit is due to both stellar and planetary tides. }
\label{fig:heating-vs-e}
\end{figure}

We analyze in Fig.~\ref{fig:heating-vs-e} how the tidal heating rates
and the orbital timescales depend on the eccentricity of the system
using all {\modif other} known parameters of the system. We first note that for values of the eccentricity $e>0.3$, these become extremely stiff functions of $e$. This implies that any initially high eccentricity value causes a rapid evolution of the system which is hardly predicted by models developed only to second order in eccentricity \citep{JGB08a,IB09,MFJ09,Gillon+10}. We find that a $10\%$ asynchronous planet would dissipate the required luminosity, but the corresponding synchronization timescale is extremely short, about $10,000$\,years.

At low eccentricities, an inward migration with a timescale $\sim 1\,{\rm Ga}\,(Q_*/10^6)$ results from tides raised by the planet onto the star.At high eccentricities and for our choice of Q factors, tides raised by the star onto the planet begin to dominate and cause a decrease in the semi-major axis that is concomitant to the circularization of stellar tides on the planet. The orbit circularization is mostly caused by the planet, unless $Q_{\rm p}> 10Q_*$. While the planet is synchronized efficiently, the star is found to be spun up by the planet relatively slowly $\sim 1\,$Ga for eccentricities $e<0.2$.

Given its inferred eccentricity, the present size of CoRoT-2b may be explained by tides only within two scenarios: (i) By a very low $Q_{\rm p}$ value and a forced eccentricity due to the presence of another planet. (ii) By an initial stage of high-dissipation followed by a rapid circularization and contraction. The former case is unlikely \citep[see also][]{Gillon+10}. The last possibility requires that the circularization proceeds faster than the planet's contraction.

\begin{figure}
\centerline{\resizebox{8.5cm}{!}{\includegraphics{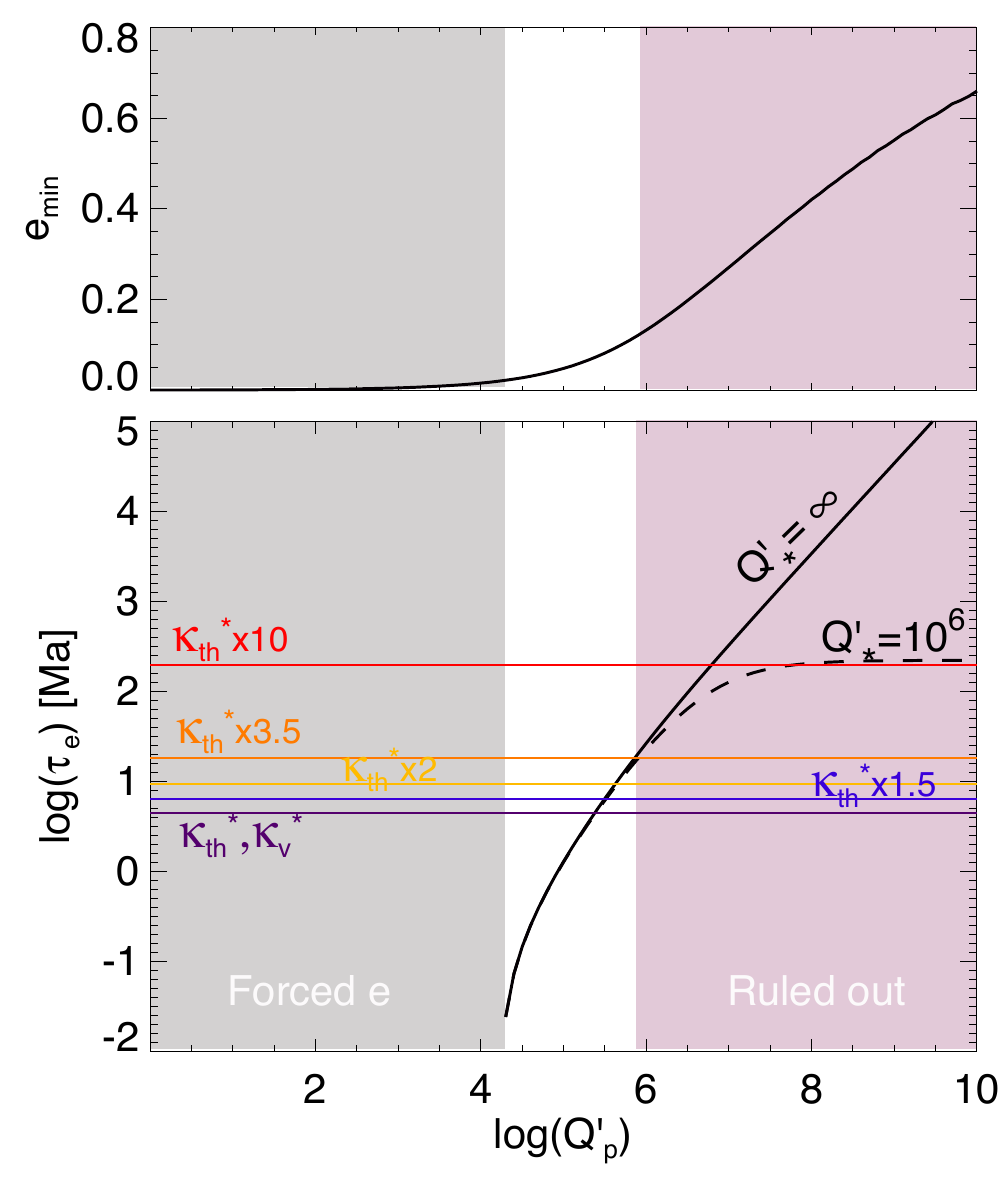}}}
\caption{Minimum eccentricity and circularization timescale as a function of tidal dissipation factor $Q'_{\rm p}$. {\em Top panel}: Eccentricity required to reach a dissipation rate $H=10^{29}\rm\,erg\,s^{-1}$ in the planet. {\em Bottom panel}: Time to evolve from the minimum eccentricity to the observed $e\sim 0.02$ for two values of the stellar tidal factor, $\infty$ (dashed) and $10^6$ (plain). The horizontal lines show the times required for CoRoT-2b to contract from $2$ to $1.5\,$\rjup, with the atmospheric boundary conditions as in Fig.~\ref{fig:atm-corot2b}.}
\label{fig:heating-qpvalues}
\end{figure}

In Fig.~\ref{fig:heating-qpvalues} we explore the constraints that can be derived on $Q_{\rm p}$. Using the equations in the Appendix, we calculate the minimum eccentricity required  for tides to dissipate $10^{29}\rm\,erg\,s^{-1}$ in the planet (top panel). We then calculate the time required for the eccentricity to decline from this value to the observed one (we assumed $e=0.02$). This time is compared to the time required to contract from $2$ to $1.5\,$\rjup\ based on our different atmospheric models (bottom panels). We thus derive an upper limit to the planetary tidal factor $Q_{\rm p}\sim 10^6$ that is similar to the $10^{5.5}$ quoted by \citet{Gillon+10}. This estimate does not account for the effect of migration but we propose that it should be relatively realistic.

\begin{figure*}
\centerline{\resizebox{14cm}{!}{\includegraphics{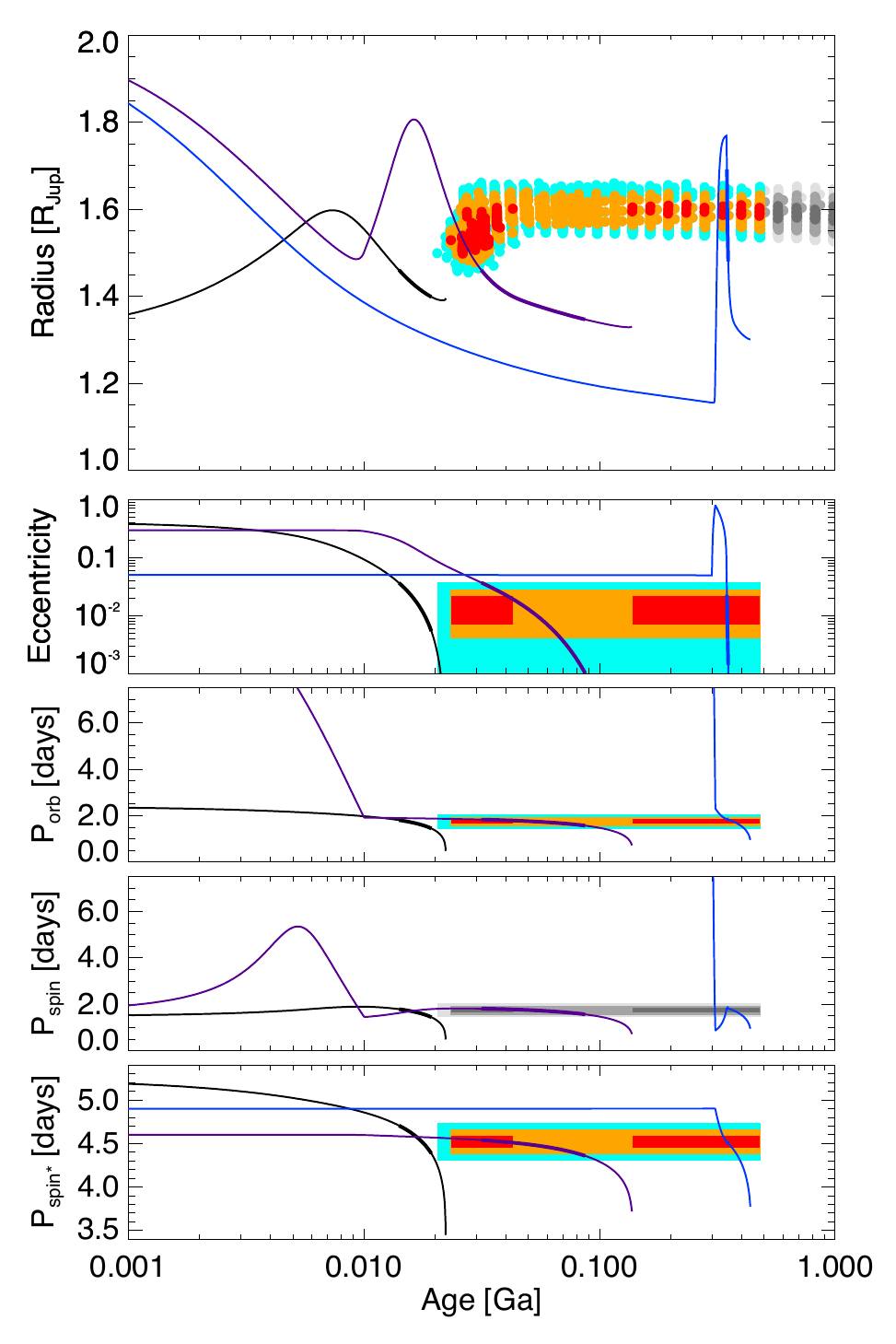}}}
\caption{Evolution of the CoRoT-2 system in the presence of tides, as a function of age expressed in billion years. From top to bottom, the five panels are: (a) Planet's transit radius; (b) Planet's orbital eccentricity; (c) Planet's orbital period in days; (d) Planet's spin period in days; (e) Star's spin period in days. The observational constraints are indicated by colored areas (see fig.~\ref{fig:evol_compare_atms}). For visibility purposes, the constraint on the orbital period has been increased by a factor $10^5$. Three models are shown: {\em Black lines}: A model with $Q'_{\rm p}=3\times 10^5$, $Q'_*=10^5$, $e_{\rm ini}=0.4$, $a_{\rm ini}=0.04\,$AU, corresponding to a preferred model of \citet{Gillon+10}. {\em Purple lines}: A model including migration for the first 10\,Ma, and $Q'_{\rm p}=10^6$, $Q'_*=10^6$, $e_{\rm ini}=0.4$, $a_{\rm ini}=0.13\,$AU.  {\em Blue lines}: A close-encounter model, for which the eccentricity is suddenly raised to $e\sim 0.77$ after 300\,Ma, presumably due to close encounters with another planet. This assumes $Q'_{\rm p}=10^5$, $Q'_*=10^6$, $a_{\rm ini}=0.1\,$AU. The thick parts of the lines corresponds to epochs for which the eccentricity and orbital periods are in agreement with the observations (see text).}
\label{fig:evol_compare_tides}
\end{figure*}

We show in Fig.~\ref{fig:evol_compare_tides} a few example results of the full dynamical calculation. One problem we faced was the existence of a runaway inflation of the planet especially for high values of the eccentricity. In order to maximize the ensemble of solutions (and ease the calculation), we set an arbitrary saturation value at $H=2\times 10^{29}\,\rm erg\,s^{-1}$ and increase the planetary $Q_{\rm p}$ factor accordingly.  Using this recipe, we are able to reproduce approximately the solutions found by \citet{Gillon+10}, but find that the contraction is too fast compared to our age constraints. We were unable in general to find solutions when starting from a fixed initial eccentricity at time $t=0$. We can obtain transient solutions instead when including inward migration or a late increase in the eccentricity. A plausible scenario could thus be that CoRoT-2b had a recent ($\wig{<} 20$\,Ma) planetary encounter leaving it into an eccentric orbit \citep[e.g.][]{JT08,FR08}. Another similar possibility is that the planet had a Kozai interaction with another distant body \citep[e.g.][]{FT07}, its orbit was changed to one of high eccentricity, and circularization began less than $20$\,Ma ago.

\subsection{A recent planetary impact?}

Another alternative is of course that CoRoT-2b experienced a planetary
impact in the past $\sim 20\,$Ma. \citet{JT08} find that up to about
$20\%$ of planets may experience mergers in systems with multiple
embryos of the same mass. This probability is extremely dependent on
initial conditions, but illustrates that giant impacts may
not be an extremely rare possibility. We illustrate the possible mass
of such an impactor with an extremely simple model. The energy
required to inflate the precursor of CoRoT-2b is 
\begin{equation}
\Delta E_{\rm grav}\sim {GM_0^2\over R_0} \beta,
\end{equation}
 where $M_0$ is the mass of the precursor, $R_0$ its size, and $\beta$ is the radius factor increase required to explain CoRoT-2b's present size. We assume that the collision takes place at the escape velocity of the precursor, and that half of the kinetic energy is transferred into increasing the internal energy of the final planet, i.e., that
\begin{equation}
\Delta E_{\rm collision}\sim {1\over 4} {GM_0 M_1 \over R_0},
\end{equation}
 where $M_1$ is the mass of the impactor. Using $M_{\rm p}\sim M_0+
 M_1$ as the present mass of CoRoT-2b, and $\Delta E_{\rm
   collision}\sim \Delta E_{\rm grav}$  we obtain
\begin{equation}
 M_1\sim {\beta\over \beta +1/4} M_{\rm p}.
\end{equation}
 This relation is independent of the initial planetary
 radius. This is because both the impact energy and the inflation
 energy have the same $R_0$ dependency.  If we assume $\beta\approx
 1/4$, we find $M_1\sim 1/2\, M_{\rm p}\sim 1.75\,$\mjup. If the
 system is really young, the need for a large impactor decreases:
 e.g. with $\beta\approx 0.05$, we obtain $M_1\sim 1/6\, M_{\rm p}\sim
 0.6\,$\mjup. These values of course represent extremely simplified
 examples but they show that unless planets can be accelerated to much higher
 velocities (this would probably require a fourth, massive high density
 object, i.e. a brown dwarf), only collisions between giant planets would provide the
 necessary energy to significantly inflate CoRoT-2b.

\section{Conclusion}

We have combined stellar and planetary evolution models to help us develop a
consistent scenario to understand the formation and evolution of
the CoRoT-2 system. Although stellar spots are important in the
lightcurve analysis and do have an important impact when deriving the
planetary radius, we have demonstrated that their consequences on the derived
stellar properties are modest and can be modeled as an additional uncertainty in the derived effective temperature of the star.

We have presented evidence for the youth of the CoRoT-2 system. The rapid $\sim 4.5$\,day spin of the star is of course a strong indication of a young age, but we have also shown that a very promising ensemble of solutions occurs on the pre-main-sequence phase of the parent star, yielding ages of between 30 and 40\,Ma.
By combining constraints obtained on the atmosphere from both brightness
temperature measurements and stellar and planetary evolution models, we
found that three scenarios can explain the present large size of planet
CoRoT-2b. All these imply a recent event that took place less than 40\,Ma ago.
\begin{enumerate}
\item Its atmosphere is about 3 to 4 times more opaque in the infrared than usually thought, and the system is indeed 30 to 40\,Ma old.
\item The CoRoT-2 system involved multiple planets and less than 20\,Ma ago, close encounters between the planets brought two into collision. The CoRoT-2b impactor would need to have had a relatively significant fraction of the total planetary mass.
\item The CoRoT-2 system involved multiple planets and less than 20\,Ma ago, close encounters between the planets left CoRoT-2b on an eccentric orbit. This orbit has been almost circularized, but remains in the transient heating phase. The parameter space required in terms of initial eccentricities and orbital distances for this to have occurred is small, and we found that it generally requires a saturation of the tidal heating to avoid the complete loss of the planet.
\end{enumerate}

{\modif The detection of abundant lithium in CoRoT-2 confirms that the
  star is young: the comparison with data obtained for open clusters
  indicate that its age should be between 30 and 320\,Ma
  \citep{Gillon+10}. However, this may be an overestimation because }
lithium may be more easily destroyed in the presence of a massive
protoplanetary disk \citep[see][]{Bouvier08}.  Several kinds of
observations of CoRoT-2 would shed light on its nature. The detection
of an infrared excess or a debris disk would be an indication that
the system is young even though there is a wide spread in disk ages
\citep[e.g.][]{Hillenbrand+08}. Searching for distant companions would
also help especially given the possibility that the planet's
eccentricity may have been pumped up by Kozai interactions with a
third body before being efficiently damped by tides. Last but not
least, the determination of a complete infrared lightcurve including
both the primary and the secondary transit would be extremely valuable
to constrain the planet's atmospheric properties, in particular its
day-night heat redistribution efficiency. {\modif It would also help us
  to determine the planet-to-star radius ratio in way that is less affected by
  systematic errors due to stellar activity.} With
similar observations in the visible, one would be able to test our assumption
that the signal detected in the CoRoT lightcurves is due to thermal
emission from deep levels in the planet. Since the radiative
timescales rapidly increase with depth \citep{IBG05}, we would expect
that phase variations in the visible are smaller than in the infrared.

\section*{Acknowledgments}

{\modif We thank an anonymous referee for helpful comments.} The
present study was made possible thanks to observations obtained with
CoRoT, a space project operated by the French Space Agency, CNES, with
participation of the Science Program of ESA, ESTEC/RSSD, Austria,
Belgium, Brazil, Germany and Spain. We acknowledge the support of the
{\em Programme National de Plan\'etologie} and of CNES. Computations
have been done on the {\em Mesocentre SIGAMM} machine, hosted by the
Observatoire de la C\^ote d'Azur.

\bibliography{corot2}

\begin{thebibliography}{57}
\expandafter\ifx\csname natexlab\endcsname\relax\def\natexlab#1{#1}\fi

\bibitem[{{Allard} {et~al.}(2001){Allard}, {Hauschildt}, {Alexander},
  {Tamanai}, \& {Schweitzer}}]{Allard+01}
{Allard}, F., {Hauschildt}, P.~H., {Alexander}, D.~R., {Tamanai}, A., \&
  {Schweitzer}, A. 2001, \apj, 556, 357

\bibitem[{{Alonso} {et~al.}(2008){Alonso}, {Auvergne}, {Baglin}, {Ollivier},
  {Moutou}, {Rouan}, {Deeg}, {Aigrain}, {Almenara}, {Barbieri}, {Barge},
  {Benz}, {Bord{\'e}}, {Bouchy}, {de La Reza}, {Deleuil}, {Dvorak}, {Erikson},
  {Fridlund}, {Gillon}, {Gondoin}, {Guillot}, {Hatzes}, {H{\'e}brard},
  {Kabath}, {Jorda}, {Lammer}, {L{\'e}ger}, {Llebaria}, {Loeillet}, {Magain},
  {Mayor}, {Mazeh}, {P{\"a}tzold}, {Pepe}, {Pont}, {Queloz}, {Rauer},
  {Shporer}, {Schneider}, {Stecklum}, {Udry}, \& {Wuchterl}}]{Alonso+08}
{Alonso}, R., {Auvergne}, M., {Baglin}, A., {et~al.} 2008, \aap, 482, L21

\bibitem[{{Alonso} {et~al.}(2010{\natexlab{a}}){Alonso}, {Deeg}, {Kabath}, \&
  {Rabus}}]{ADKR10}
{Alonso}, R., {Deeg}, H.~J., {Kabath}, P., \& {Rabus}, M. 2010{\natexlab{a}},
  \aj, 139, 1481

\bibitem[{{Alonso} {et~al.}(2009){Alonso}, {Guillot}, {Mazeh}, {Aigrain},
  {Alapini}, {Barge}, {Hatzes}, \& {Pont}}]{Alonso+09}
{Alonso}, R., {Guillot}, T., {Mazeh}, T., {et~al.} 2009, \aap, 501, L23

\bibitem[{{Alonso} {et~al.}(2010{\natexlab{b}}){Alonso}, {Guillot}, {Mazeh},
  {Aigrain}, {Alapini}, {Barge}, {Hatzes}, \& {Pont}}]{Alonso+10}
{Alonso}, R., {Guillot}, T., {Mazeh}, T., {et~al.} 2010{\natexlab{b}}, \aap, in
  press

\bibitem[{{Ammler-von Eiff} {et~al.}(2009){Ammler-von Eiff}, {Santos}, {Sousa},
  {Fernandes}, {Guillot}, {Israelian}, {Mayor}, \& {Melo}}]{AmmlervonEiff+09}
{Ammler-von Eiff}, M., {Santos}, N.~C., {Sousa}, S.~G., {et~al.} 2009, \aap,
  507, 523

\bibitem[{{Baraffe} {et~al.}(1998){Baraffe}, {Chabrier}, {Allard}, \&
  {Hauschildt}}]{BCAH98}
{Baraffe}, I., {Chabrier}, G., {Allard}, F., \& {Hauschildt}, P.~H. 1998, \aap,
  337, 403

\bibitem[{{Baraffe} {et~al.}(2003){Baraffe}, {Chabrier}, {Barman}, {Allard}, \&
  {Hauschildt}}]{Baraffe+03}
{Baraffe}, I., {Chabrier}, G., {Barman}, T.~S., {Allard}, F., \& {Hauschildt},
  P.~H. 2003, \aap, 402, 701

\bibitem[{{Barker} \& {Ogilvie}(2009)}]{BO09}
{Barker}, A.~J. \& {Ogilvie}, G.~I. 2009, \mnras, 395, 2268

\bibitem[{Beatty {et~al.}(2007)Beatty, Fernandez, Latham, Bakos, Kovacs, Noyes,
  Stefanik, Torres, Everett, \& Hergenrother}]{Beatty2007}
Beatty, T.~G., Fernandez, J.~M., Latham, D.~W., {et~al.} 2007, \apj, 663, 573

\bibitem[{{Bodenheimer} {et~al.}(2001){Bodenheimer}, {Lin}, \&
  {Mardling}}]{BLM01}
{Bodenheimer}, P., {Lin}, D.~N.~C., \& {Mardling}, R.~A. 2001, \apj, 548, 466

\bibitem[{{Bouchy} {et~al.}(2008){Bouchy}, {Queloz}, {Deleuil}, {Loeillet},
  {Hatzes}, {Aigrain}, {Alonso}, {Auvergne}, {Baglin}, {Barge}, {Benz},
  {Bord{\'e}}, {Deeg}, {de La Reza}, {Dvorak}, {Erikson}, {Fridlund},
  {Gondoin}, {Guillot}, {H{\'e}brard}, {Jorda}, {Lammer}, {L{\'e}ger},
  {Llebaria}, {Magain}, {Mayor}, {Moutou}, {Ollivier}, {P{\"a}tzold}, {Pepe},
  {Pont}, {Rauer}, {Rouan}, {Schneider}, {Triaud}, {Udry}, \&
  {Wuchterl}}]{Bouchy+08}
{Bouchy}, F., {Queloz}, D., {Deleuil}, M., {et~al.} 2008, \aap, 482, L25

\bibitem[{{Bouvier}(2008)}]{Bouvier08}
{Bouvier}, J. 2008, \aap, 489, L53

\bibitem[{{Burrows} {et~al.}(2007){Burrows}, {Hubeny}, {Budaj}, \&
  {Hubbard}}]{BHBH07}
{Burrows}, A., {Hubeny}, I., {Budaj}, J., \& {Hubbard}, W.~B. 2007, \apj, 661,
  502

\bibitem[{{Czesla} {et~al.}(2009){Czesla}, {Huber}, {Wolter}, {Schr{\"o}ter},
  \& {Schmitt}}]{Czesla+09}
{Czesla}, S., {Huber}, K.~F., {Wolter}, U., {Schr{\"o}ter}, S., \& {Schmitt},
  J.~H.~M.~M. 2009, ArXiv e-prints

\bibitem[{{Demarque} {et~al.}(2004){Demarque}, {Woo}, {Kim}, \&
  {Yi}}]{Demarque+04}
{Demarque}, P., {Woo}, J., {Kim}, Y., \& {Yi}, S.~K. 2004, \apjs, 155, 667

\bibitem[{{Eggleton} {et~al.}(1998){Eggleton}, {Kiseleva}, \&
  {Hut}}]{Eggleton98}
{Eggleton}, P.~P., {Kiseleva}, L.~G., \& {Hut}, P. 1998, \apj, 499, 853

\bibitem[{{Fabrycky} \& {Tremaine}(2007)}]{FT07}
{Fabrycky}, D. \& {Tremaine}, S. 2007, \apj, 669, 1298

\bibitem[{{Ford} \& {Rasio}(2008)}]{FR08}
{Ford}, E.~B. \& {Rasio}, F.~A. 2008, \apj, 686, 621

\bibitem[{{Fortney} {et~al.}(2008){Fortney}, {Lodders}, {Marley}, \&
  {Freedman}}]{FLMF08}
{Fortney}, J.~J., {Lodders}, K., {Marley}, M.~S., \& {Freedman}, R.~S. 2008,
  \apj, 678, 1419

\bibitem[{{Fr{\"o}hlich} \& {Lean}(2004)}]{FL04}
{Fr{\"o}hlich}, C. \& {Lean}, J. 2004, \aapr, 12, 273

\bibitem[{{Gillon} {et~al.}(2010){Gillon}, {Lanotte}, {Barman}, {Miller},
  {Demory}, {Deleuil}, {Montalb{\'a}n}, {Bouchy}, {Collier Cameron}, {Deeg},
  {Fortney}, {Fridlund}, {Harrington}, {Magain}, {Moutou}, {Queloz}, {Rauer},
  {Rouan}, \& {Schneider}}]{Gillon+10}
{Gillon}, M., {Lanotte}, A.~A., {Barman}, T., {et~al.} 2010, \aap, 511, A3+

\bibitem[{{Grevesse} \& {Noels}(1993)}]{Grevesse+93}
{Grevesse}, N. \& {Noels}, A. 1993, Physica Scripta Volume T, 47, 133

\bibitem[{{Gu} {et~al.}(2003){Gu}, {Lin}, \& {Bodenheimer}}]{GLB03}
{Gu}, P., {Lin}, D.~N.~C., \& {Bodenheimer}, P.~H. 2003, \apj, 588, 509

\bibitem[{{Guillot}(2005)}]{Guillot05}
{Guillot}, T. 2005, Annual Review of Earth and Planetary Sciences, 33, 493

\bibitem[{{Guillot}(2008)}]{Guillot08}
{Guillot}, T. 2008, Physica Scripta Volume T, 130, 014023

\bibitem[{{Guillot}(2010)}]{Guillot10}
{Guillot}, T. 2010, submitted to \aap

\bibitem[{{Guillot} \& {Morel}(1995)}]{GM95}
{Guillot}, T. \& {Morel}, P. 1995, \aaps, 109, 109

\bibitem[{{Guillot} {et~al.}(2006){Guillot}, {Santos}, {Pont}, {Iro}, {Melo},
  \& {Ribas}}]{Guillot+06}
{Guillot}, T., {Santos}, N.~C., {Pont}, F., {et~al.} 2006, \aap, 453, L21

\bibitem[{{Guillot} \& {Showman}(2002)}]{GS02}
{Guillot}, T. \& {Showman}, A.~P. 2002, \aap, 385, 156

\bibitem[{{Gustafsson} {et~al.}(2008){Gustafsson}, {Edvardsson}, {Eriksson},
  {J{\o}rgensen}, {Nordlund}, \& {Plez}}]{Gustafsson+08}
{Gustafsson}, B., {Edvardsson}, B., {Eriksson}, K., {et~al.} 2008, \aap, 486,
  951

\bibitem[{{Hansen}(2008)}]{Hansen08}
{Hansen}, B.~M.~S. 2008, \apjs, 179, 484

\bibitem[{{Hillenbrand} {et~al.}(2008){Hillenbrand}, {Carpenter}, {Kim},
  {Meyer}, {Backman}, {Moro-Mart{\'{\i}}n}, {Hollenbach}, {Hines}, {Pascucci},
  \& {Bouwman}}]{Hillenbrand+08}
{Hillenbrand}, L.~A., {Carpenter}, J.~M., {Kim}, J.~S., {et~al.} 2008, \apj,
  677, 630

\bibitem[{{Huber} {et~al.}(2010){Huber}, {Czesla}, {Wolter}, \&
  {Schmitt}}]{HCWS10}
{Huber}, K.~F., {Czesla}, S., {Wolter}, U., \& {Schmitt}, J.~H.~M.~M. 2010,
  ArXiv e-prints

\bibitem[{{Hut}(1981)}]{Hut81}
{Hut}, P. 1981, \aap, 99, 126

\bibitem[{{Ibgui} \& {Burrows}(2009)}]{IB09}
{Ibgui}, L. \& {Burrows}, A. 2009, \apj, 700, 1921

\bibitem[{{Iro} {et~al.}(2005){Iro}, {B{\'e}zard}, \& {Guillot}}]{IBG05}
{Iro}, N., {B{\'e}zard}, B., \& {Guillot}, T. 2005, \aap, 436, 719

\bibitem[{{Jackson} {et~al.}(2008){Jackson}, {Greenberg}, \& {Barnes}}]{JGB08a}
{Jackson}, B., {Greenberg}, R., \& {Barnes}, R. 2008, \apj, 678, 1396

\bibitem[{{Juri{\'c}} \& {Tremaine}(2008)}]{JT08}
{Juri{\'c}}, M. \& {Tremaine}, S. 2008, \apj, 686, 603

\bibitem[{{Krivova} \& {Solanki}(2008)}]{KS08}
{Krivova}, N.~A. \& {Solanki}, S.~K. 2008, Journal of Astrophysics and
  Astronomy, 29, 151

\bibitem[{{Krivova} {et~al.}(2006){Krivova}, {Solanki}, \&
  {Floyd}}]{Krivova+06}
{Krivova}, N.~A., {Solanki}, S.~K., \& {Floyd}, L. 2006, \aap, 452, 631

\bibitem[{{Lanza} {et~al.}(2009){Lanza}, {Pagano}, {Leto}, {Messina},
  {Aigrain}, {Alonso}, {Auvergne}, {Baglin}, {Barge}, {Bonomo}, {Boumier},
  {Collier Cameron}, {Comparato}, {Cutispoto}, {de Medeiros}, {Foing},
  {Kaiser}, {Moutou}, {Parihar}, {Silva-Valio}, \& {Weiss}}]{Lanza+09}
{Lanza}, A.~F., {Pagano}, I., {Leto}, G., {et~al.} 2009, \aap, 493, 193

\bibitem[{{Leconte} {et~al.}(2009){Leconte}, {Baraffe}, {Chabrier}, {Barman},
  \& {Levrard}}]{Leconte+09}
{Leconte}, J., {Baraffe}, I., {Chabrier}, G., {Barman}, T., \& {Levrard}, B.
  2009, \aap, 506, 385

\bibitem[{{Leconte} {et~al.}(2010){Leconte}, {Chabrier}, {Baraffe}, \&
  {Levrard}}]{LCBL10}
{Leconte}, J., {Chabrier}, G., {Baraffe}, I., \& {Levrard}, B. 2010, ArXiv
  e-prints

\bibitem[{{Lubow} {et~al.}(1997){Lubow}, {Tout}, \& {Livio}}]{LTL97}
{Lubow}, S.~H., {Tout}, C.~A., \& {Livio}, M. 1997, \apj, 484, 866

\bibitem[{{Mamajek} \& {Hillenbrand}(2008)}]{MH08}
{Mamajek}, E.~E. \& {Hillenbrand}, L.~A. 2008, \apj, 687, 1264

\bibitem[{{Miller} {et~al.}(2009){Miller}, {Fortney}, \& {Jackson}}]{MFJ09}
{Miller}, N., {Fortney}, J.~J., \& {Jackson}, B. 2009, \apj, 702, 1413

\bibitem[{{Morel} \& {Lebreton}(2008)}]{ML08}
{Morel}, P. \& {Lebreton}, Y. 2008, \apss, 316, 61

\bibitem[{{Ogilvie} \& {Lin}(2004)}]{OL04}
{Ogilvie}, G.~I. \& {Lin}, D.~N.~C. 2004, \apj, 610, 477

\bibitem[{{Rowe} {et~al.}(2008){Rowe}, {Matthews}, {Seager}, {Miller-Ricci},
  {Sasselov}, {Kuschnig}, {Guenther}, {Moffat}, {Rucinski}, {Walker}, \&
  {Weiss}}]{Rowe+08}
{Rowe}, J.~F., {Matthews}, J.~M., {Seager}, S., {et~al.} 2008, \apj, 689, 1345

\bibitem[{{Saumon} {et~al.}(1995){Saumon}, {Chabrier}, \& {van Horn}}]{SCvH95}
{Saumon}, D., {Chabrier}, G., \& {van Horn}, H.~M. 1995, \apjs, 99, 713

\bibitem[{{Showman} \& {Guillot}(2002)}]{SG02}
{Showman}, A.~P. \& {Guillot}, T. 2002, \aap, 385, 166

\bibitem[{{Silva-Valio} {et~al.}(2010){Silva-Valio}, {Lanza}, {Alonso}, \&
  {Barge}}]{Silva-Valio+10}
{Silva-Valio}, A., {Lanza}, A.~F., {Alonso}, R., \& {Barge}, P. 2010, \aap,
  510, A25+

\bibitem[{{Snellen} {et~al.}(2010){Snellen}, {de Mooij}, \&
  {Burrows}}]{Snellen+10}
{Snellen}, I.~A.~G., {de Mooij}, E.~J.~W., \& {Burrows}, A. 2010, \aap, 513,
  A76+

\bibitem[{{Solanki} \& {Fligge}(2000)}]{SF00}
{Solanki}, S.~K. \& {Fligge}, M. 2000, Space Science Reviews, 94, 127

\bibitem[{Sozzetti {et~al.}(2007)Sozzetti, Torres, Charbonneau, Latham, Holman,
  Winn, Laird, \& O’Donovan}]{Sozzetti2007}
Sozzetti, A., Torres, G., Charbonneau, D., {et~al.} 2007, \apj, 664, 1190

\bibitem[{{Tingley} \& {Sackett}(2005)}]{Tingley05}
{Tingley}, B. \& {Sackett}, P.~D. 2005, \apj, 627, 1011

\end{thebibliography}

\section*{Appendix}

We present the time-averaged equations for the dynamical evolution of
the star-planet system used in this work. These were taken from
\citet{BO09} and applied to the case of a planar system
(i.e. neglecting possible inclinations of the star and planet),
accounting for external forcing of the semi-major axis, eccentricity,
and the conservation of angular momentum during the planet's
contraction.

Given a semi-major axis $a$ and an eccentricy $e$, the orbital mean
motion is $n=(G(m_1+m_2)/a^3)^{1/2}$ and the angular momentum
$h=na^2(1-e^2)^{1/2}$. The equations of secular evolution of $h$, $e$,
the star's spin $\Omega_1$ and planet's spin $\Omega_2$ under the
effect of tides are:
\begin{eqnarray}
  {1\over h}{d h\over dt}&=&-{1\over t_{f1}}\left[-{\Omega_1\over 2n}f_3(e^2)+\left(f_4(e^2)-{\Omega_1\over 2n}f_2{e^2}\right)\right] \nonumber\\
  &&-{1\over t_{f2}}\left[-{\Omega_2\over 2n}f_3(e^2)+\left(f_4(e^2)-{\Omega_2\over 2n}f_2{e^2}\right)\right]\nonumber\\
  &&-{e\over 1-e^2}\dot{e}+{1\over 2a}\dot{a},\label{eq:dhdt}\\
{de\over dt} &=& -{1\over t_{f1}}\left[9\left(f_1(e^2)-{11\over 18}{\Omega_1\over n}f_2(e^2)\right)e\right]\nonumber\\
 && -{1\over t_{f2}}\left[9\left(f_1(e^2)-{11\over 18}{\Omega_2\over n}f_2(e^2)\right)e\right]+\dot{e},\label{eq:dedt}\\
{d\Omega_1\over dt}&=&{\mu\over I_1t_{f1}}\left[-{\Omega_1\over 2n}\left(f_3(e^2)+f_2(e^2)\right)+f_4(e^2)\right]h+\dot{\Omega}_1 ,\label{eq:do1dt}\\
{d\Omega_2\over dt}&=&{\mu\over I_2t_{f2}}\left[-{\Omega_2\over 2n}\left(f_3(e^2)+f_2(e^2)\right)+f_4(e^2)\right]h+\dot{\Omega}_2 .\label{eq:do2dt}
\end{eqnarray}

In these equations, $\mu=m_1 m_2/(m_1+m_2)$ is the reduced mass of the system, $I_{1,2}$ are the moments of inertia of the star and planet, and $t_{f1,2}$ are the tidal friction timescales, defined as
\begin{eqnarray}
{1\over t_{f1}}&=&\left(9n\over 2Q'_1\right)\left(m_2\over m_1\right)\left(R_1\over a\right)^5,\\
{1\over t_{f2}}&=&\left(9n\over 2Q'_2\right)\left(m_1\over m_2\right)\left(R_2\over a\right)^5,
\end{eqnarray}
where $Q'_{1,2}$ are tidal dissipation efficiency parameters defined as $Q'_{1,2}=3/(2k_{1,2}n\tau_{1,2})$, $k_{1,2}$ being the second-order Love numbers for the star and planet, and $\tau_{1,2}$ the assumed constant lag time between the quasi-hydrostatic figure and its equilibrium tide value \citep[see][for a discussion]{BO09,Eggleton98}.

The quantities $\dot{a}$, $\dot{e}$, $\dot{\Omega}_1$, and $\dot{\Omega}_2$ correspond to imposed rates of change per unit time of the semi-major axis, eccentricity, star spin, and planet spin, respectively. They are all set to mimic the presence of additional physical processes: $\dot{a}\ne 0$ accounts for the migration imposed by a circumstellar disk on the planet, $\dot{e}\ne 0$ mimics the mean eccentricity increase due to a Kozai effect, and $\dot{\Omega}_1\ne 0$ accounts for the evolution of the stellar spin due to magnetic breaking. The rate of change of the planet spin $\dot{\Omega}_2$ however, is self-consistently obtained from angular momentum conservation considerations that account for the planet's contraction rate $\dot{R}_2$ obtained from the evolution calculations
\begin{equation}
\dot{\Omega}_2=-2{\Omega_2\over R_2}\dot{R}_2.
\label{eq:omega_2}
\end{equation}
This relation assumes that the planet rotates as a solid body, and that its moment of inertia remains constant.

Furthermore, Eqs~(\ref{eq:dhdt}-\ref{eq:do2dt}) make use of the following functions of the eccentricity \citep{BO09, Hut81}:
\begin{eqnarray}
f_1(e^2)&=&\left(1+{15\over 4}e^2+{15\over 8}e^4+{5\over 64}e^6\right)\left(1-e^2\right)^{-13/2}\\
f_2(e^2)&=&\left(1+{3\over 2}e^2+{1\over 8}e^4\right)\left(1-e^2\right)^{-5}\\
f_3(e^2)&=&\left(1+{9\over 2}e^2+{5\over 8}e^4\right)\left(1-e^2\right)^{-5}\\
f_4(e^2)&=&\left(1+{15\over 2}e^2+{45\over 8}e^4+{5\over 16}e^6\right)\left(1-e^2\right)^{-13/2}\\
f_5(e^2)&=&\left(3+{1\over 2}e^2\right)\left(1-e^2\right)^{-5}\\
f_6(e^2)&=&\left(1+{31\over 2}e^2+{255\over 8}e^4+{185\over 16}e^6+{25\over 64}\right)\left(1-e^2\right)^{-8}
\end{eqnarray}

Last but not least, from energy conservation, one infers that the energy dissipated in the star and in the planet, respectively, are
\begin{eqnarray}
H_1=\mu{h\over n t_{f1}}&&\left[{1\over 2}\Omega_1^2\left(f_3(e^2)+f_2(e^2)\right)-2n\Omega_1f_4(e^2)\right.\nonumber\\
&&\left.+n^2f_6(e^2)\right],\label{eq:H1}\\
H_2=\mu{h\over n t_{f2}}&&\left[{1\over 2}\Omega_2^2\left(f_3(e^2)+f_2(e^2)\right)-2n\Omega_2f_4(e^2)\right.\nonumber\\
&&\left.+n^2f_6(e^2)\right].\label{eq:H2}
\end{eqnarray}
The heating rate in the planet $H_2$ may then be applied to the planetary evolution calculations. (Note that Eqs.~(\ref{eq:H1}) and (\ref{eq:H2}) have opposite signs compared to \citet{BO09}, so that $H_1$ and $H_2$ are positive).

For models in which tidal dissipation in the planet is assumed to saturate at a fixed value $H_2=H_{\rm 2,max}$, we proceed as follows: when the dissipation in the planet is found to exceed $H_{\rm 2,max}$, we increase the value of $Q'_2$ until we reach the maximal allowed dissipation.

\end{document}